\documentclass[msmath,amssymb,aps,pra,twocolumn,epsfig,showpacs,bibliography,lengthcheck,superscriptaddress]{revtex4-2}

\usepackage{graphicx}% Include figure files
\usepackage{dcolumn}% Align table columns on decimal point
\usepackage{bm}% bold math
\usepackage{hyperref}% add hypertext capabilities
\usepackage{epstopdf}
\usepackage{subcaption}
\usepackage{caption}
\usepackage{footmisc}
\usepackage{amsmath}
\usepackage{color}

\begin{document}
\title{Partial measurements of the total field gradient and the field gradient tensor using an atomic magnetic gradiometer}
\author{Q.-Q. Yu}
\author{S.-Q. Liu}
\author{X.-K. Wang}
\affiliation{Department of Precision Machinery and Precision Instrumentation, Key Laboratory of Precision Scientific Instrumentation of Anhui Higher Education Institutes, University of Science and Technology of China, Hefei 230027, China}
\author{D. Sheng}
\email{dsheng@ustc.edu.cn}
\affiliation{Department of Precision Machinery and Precision Instrumentation, Key Laboratory of Precision Scientific Instrumentation of Anhui Higher Education Institutes, University of Science and Technology of China, Hefei 230027, China}
\affiliation{Hefei National Laboratory, University of Science and Technology of China, Hefei 230088, China}

\begin{abstract}
Magnetic gradiometers have wide practical and academic applications, and two important types of field gradient observables are the total field gradient and field gradient tensor. However, measurements of the field gradient tensor have not been the focus of previous researches on atomic magnetic gradiometers. In this work, we develop an atomic magnetic gradiometer  based on two separately optically pumped atomic ensembles in a Herriott-cavity-assisted atomic cell. This gradiometer shows versatile operation modes and functions, and we demonstrate them in measurements of both types of field gradient observables.
\end{abstract}
\maketitle

\section{Introduction}

In the outdoor searches of objectives that are magnetic or magnetizable, magnetic field gradients usually originate from the targets while the background field gradient is normally negligible~\cite{schmidt2006}. Therefore, the field gradients contain valuable information about the targets. Moreover, with the help of differential detections, the measured field gradient is largely free of the noises and drifts in the background field. In addition, the field gradient is relatively less sensitive to the sensor orientation compared with direct vector-field measurements~\cite{getmag2004}. For these reasons, systems measuring the field gradients, magnetic gradiometers, have been widely used in underwater explorations~\cite{hirota1997}, geological surveys ~\cite{wickerham1954,overton1981,getmag2004,clark2012}, space science~\cite{acuna2002}, magnetic anomaly navigations~\cite{canciani2016}, and bio-imaging~\cite{zhangrui2020,limes2020} applications.

There are two types of field gradients that can be measured by magnetic gradiometers. One is the total field gradient, $\nabla B$, where $B$ is the total magnitude of the field. The other one is the field gradient tensor, $\nabla \mathbf{B}$, with $\mathbf{B}=\sum_i B_i\hat{r}_i$. According to basic electromagnetic relations, there are five independent elements in the field gradient tensor in free space.  Though both types of field gradients share the same aforementioned advantages in different applications, the field gradient tensor contains richer information, and the measurement results of the field gradient tensor are more convenient for geophysical interpretations~\cite{schmidt2006,stolz2006}.

Currently, practical magnetic gradiometers are mainly based on superconducting quantum interference devices (SQUIDs) and fluxgates, both of which can measure vector components of the bias field. People started to pay attention to the field gradient tensor since 1970s~\cite{wynn1975,clem1996}, and there have been great developments in this field over the past two decades~\cite{getmag2004,keenan2022,wiegert2007,pang2014,yin2018,stolz2021,stolz2022}. However, each type of sensors has its own problems. While the SQUID-based gradiometers are relatively bulky and heavy due to the additional dewar to maintain the low temperature environment, the fluxgate-based gradiometers are not sensitive and stable enough.

Atomic magnetometers are characterized by the high field sensitivity~\cite{budker2013} and potential for miniaturization~\cite{kitching2018}, which make them promising for magnetic gradiometer applications. When the bias field is low, atomic magnetometers can work in the spin-exchange-relaxation-free (SERF) regime. Since the SERF magnetometers have abilities to detect vector components of the field, gradiometers based on this technology can measure elements of the gradient tensor~\cite{sheng17,alem2017,sulai2019,kernel2021}. In the more common cases where the bias field is on the order of 10~$\mu$T, atomic magnetometers usually work in the scalar mode, and the corresponding gradiometers have only been demonstrated to detect elements of the total field gradient~\cite{zhangrui2020,limes2020,lucivero2021,perry2020}.

In this work, we develop an atomic magnetic gradiometer based on two separated $^{85}$Rb Bell-Bloom magnetometers in a single multipass cell, and demonstrate that this sensor can be used in multiple ways to measure both the total field gradient and gradient filed tensor elements in the presence of a bias field. Following this introduction, Sec.~II introduces theoretical background of this work, Sec.~III describes the Bell-Bloom scalar magnetometry and two modes of the gradiometer to partially measure the total field gradient, Sec.~IV describes two modes of the gradiometer measuring multiple elements of the field gradient tensor, and Sec.~IV concludes the paper.

\section{Theoretical background}
\subsection{Notations of the field gradients}
Considering that the background field is $\mathbf{B}_0$ and the field from the target is $\mathbf{B}_a$, the total field difference between the cases with and without the target is:
\begin{equation}
\Delta B_s=|\mathbf{B}_0+\mathbf{B}_a|-B_0,
\end{equation}
and the vector field difference is $\Delta \mathbf{B}_v=\mathbf{B}_a$. A key difference between these two parameters is that $\Delta B_v$ satisfies the Laplace equation, while it is not always true for $\Delta B_s$~\cite{schmidt2006}.

In cases that $B_a\ll B_0$, $\Delta B_s\approx \hat{\mathbf{B}}_0\cdot\mathbf{B}_a$, with $\hat{\mathbf{B}}_0$ as the unit vector of $\mathbf{B}_0$. Then the total field gradient is $\boldsymbol{G}=\nabla(\Delta B_s)$, with the component of $\boldsymbol{G}$ along the $r_i$ axis as
\begin{equation}
G_i=\frac{\partial \Delta B_s}{\partial r_i}\approx\frac{\partial{(\hat{\mathbf{B}}_0\cdot\mathbf{B}_a)}}{\partial r_i}=\sum_j\hat{\mathbf{B}}_0\cdot\hat{r}_jB_{a,ji},
\end{equation}
where $B_{a,ji}=\partial{B_{a,j}}/\partial{r_i}$ is an element of the gradient tensor $\mathbf{B}_a$. According to Maxwell equations, in free space there are only five independent gradient tensor elements of $\mathbf{B}_a$: $B_{a,xx}$, $B_{a,yy}$, $B_{a,xy}$, $B_{a,xz}$, and $B_{a,zy}$. The analytic functions formed by $G_i$ are useful to quantitatively interpret the potential-field data in two-dimensional~\cite{Nabighian1972,Nabighian1972} and three-dimensional spaces~\cite{Nabighian1984,roest1992}. More sophisticated relations between $B_{a,ij}$ and $G_i$ can be found using Fourier and Hilbert transformations~\cite{nelson1988}.

\subsection{Bell-Bloom optical pumping}
Bell-Bloom optical pumping~\cite{bell1961} is an efficient method to directly generate transverse atomic polarization. The dynamics of the electron spin polarization $\boldsymbol{P}$ is described by the Bloch equation~\cite{appelt98,shah2009}:
\begin{equation}~\label{eq:bloch}
\frac{d\boldsymbol{P}}{dt}=\gamma\boldsymbol{P}\times\boldsymbol{B}+\frac{1}{Q(P)}[R_{OP}(\boldsymbol{s}-\boldsymbol{P})-R_d\boldsymbol{P}],
\end{equation}
where $\gamma$ is the atomic gyromagnetic ratio, $Q(P)$ is the nuclear spin slowing down factor, $R_{OP}$ is the optical pumping rate, $\boldsymbol{s}$ is the photon spin of the light beam, $R_d$ is the atom depolarization rate in the absence of light. For a pump beam which is amplitude modulated at a frequency $\omega$, the optical pumping rate can be expressed as:
\begin{equation}~\label{eq:Rop}
R_{OP}=a_0+\sum_{n=1}^\infty a_n\cos(n\omega t-\alpha_n),
\end{equation}
where $a_i$ is the corresponding coefficient of the Fourier expansion series of $R_{OP}$. When the pump beam is on resonance with the Rb D1 transition and $\omega$ is close to the atomic Larmor precession frequency $\omega_L$, a substantial transverse atomic polarization can be built.

For a special configuration that the bias field is along the $z$ axis, and the pump beam propagates along the $x$ direction, the component of the atomic polarization modulated at $\omega$ can be expressed as~\cite{cai2020,yu2022}
\begin{equation}~\label{eq:pt}
P_{x}+iP_{y}=\frac{sa_1}{2[R+iQ(P)(\omega_L-\omega)]}e^{-i(\omega t-\alpha_1)},
\end{equation}
where $R=R_d+a_0$. If the a linearly polarized off-resonant probe beam propagates along the $y$ axis, the rotation of the probe beam polarization due to the photon-atom interaction is proportional to the real part of $P_y$ in Eq.~\eqref{eq:pt}~\cite{happer72},
\begin{eqnarray}~\label{eq:Py}
Re(P_y)=\frac{sa_1}{2}\left[\frac{-R\sin(\omega t-\alpha_1)}{R^2+Q^2(P)(\omega_L-\omega)^2}\right.-\nonumber\\
\left.\frac{Q(P)(\omega_L-\omega)\cos(\omega t-\alpha_1)}{R^2+Q^2(P)(\omega_L-\omega)^2}\right..
\end{eqnarray}

For a more general case, the bias field is along an arbitrary direction, so that the angle between the bias field and the pump beam direction ($x$ axis) is $\psi_x$, and the angle between the projection of the bias field on $y$-$z$ plane and the $y$ axis is $\theta_{yz}$. The component of the atomic polarization modulated at $\omega$ along the $y$ axis is~\cite{cai2020}:
\begin{eqnarray}~\label{eq:Py2}
P_y=&&\frac{sa_1d\sin\psi_x}{2}\left[\frac{-R\cos(\omega t-\alpha_1-\beta)}{R^2+Q^2(P)(\omega_L-\omega)^2}\right.\nonumber\\
&&+\left.\frac{Q(P)(\omega_L-\omega)\sin(\omega t-\alpha_1-\beta)}{R^2+Q^2(P)(\omega_L-\omega)^2}\right],
\end{eqnarray}
where $d=\sqrt{\cos^2\psi_x\cos^2\theta_{yz}+\sin^2\theta_{yz}}$, and $\beta=\sin^{-1}(\sin\theta_{yz}/d)$.

Therefore, if the driving signal for modulating the pump beam amplitude is used to demodulate the signal due to the probe beam polarization rotation, we will get Lorentzian and  dispersion line shapes in the demodulation outputs with suitable phase choices.

\subsection{Effect of a half-wave plate on the probe beam}~\label{sec:hfplate}
For a linearly polarized probe beam with the polarization tilted from the $x$ axis by an angle of $\theta$, its polarization can be expressed by the Jones vector as $P_i=(\cos\theta, \sin\theta)$~\cite{jones1941}. Suppose that it passes two polarized atomic ensembles successively, and the polarization of the probe beam is rotated by $\alpha$ and $\beta$ due to the interaction between photon and atoms in each ensemble, respectively. Then, the polarization of the transmitted probe beam is
\begin{equation}
P_o=R(\beta)R(\alpha)P_i=(\cos(\theta+\alpha+\beta),\sin(\theta+\alpha+\beta)),
\end{equation}
where $R$ is the two-dimensional rotation matrix.

If we put a half-wave plate (fast axis along the $x$ axis) in between the two atomic ensembles, then the  polarization of the transmitted probe beam is modified to
\begin{equation}
P_o=R(\beta)H R(\alpha)P_i=(\cos(\theta+\alpha-\beta),\sin(\theta+\alpha-\beta)),
\end{equation}
where $H$ is the Jones matrix for the half-wave plate. It can be concluded from the equation above that the probe beam differentially detects the two polarized atomic ensembles separated by this half-wave plate.

\section{Partial measurement of the total field gradient}

\begin{figure}[htb]
\includegraphics[width=3in]{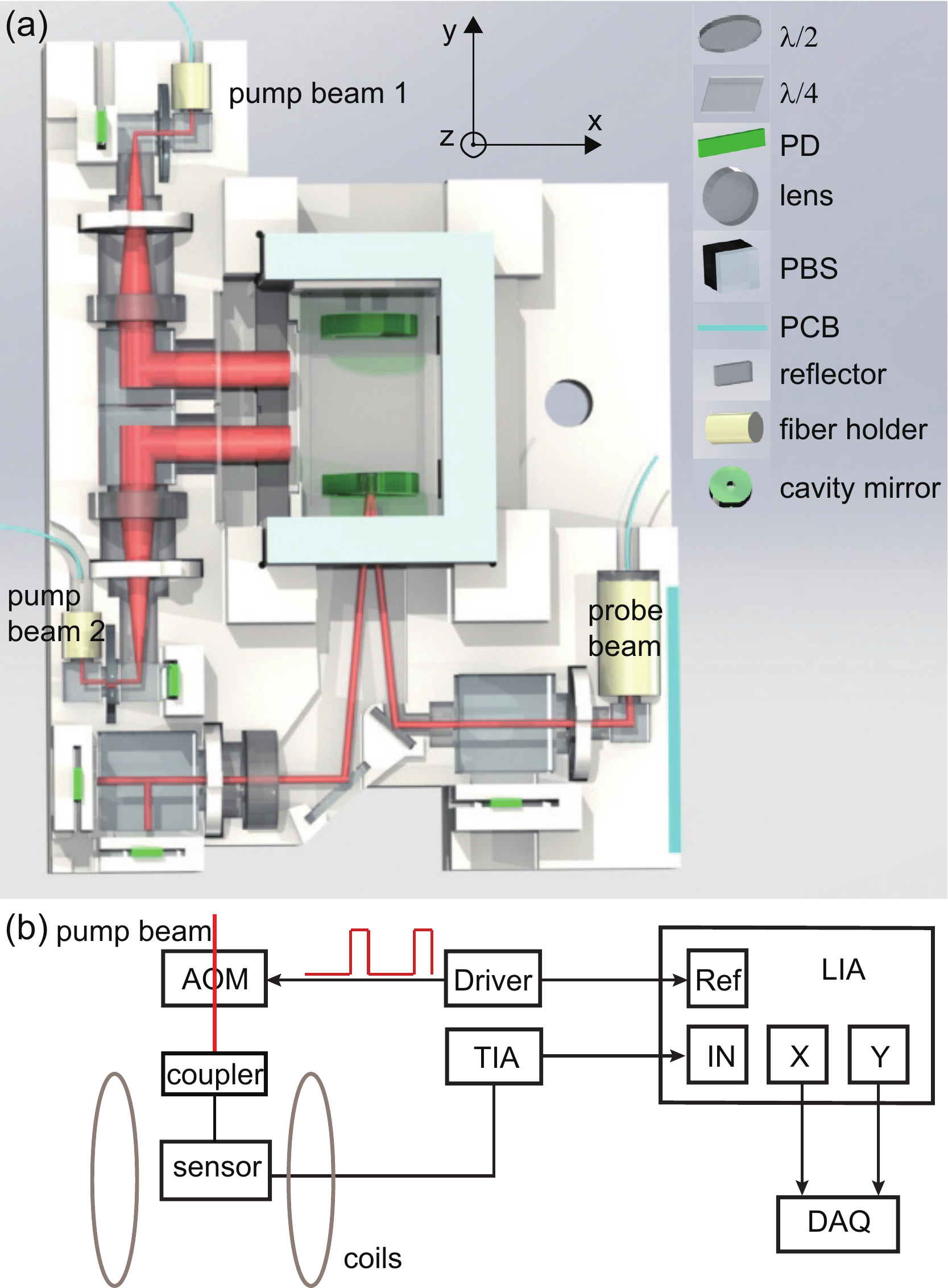}
\caption{\label{fig:setup} Plot~(a) illustrates the gradient magnetometer. $\lambda/2$: half-wave plate, $\lambda/4$: quarter-wave plate. Plot~(b) shows the electrical signal processing of a Bell-Bloom scalar magnetometer. DAQ: data acquisition.}
\end{figure}

The sensor used in this work is shown in Fig.~\ref{fig:setup}(a). A key element of this sensor is a Herriott-cavity-assisted vapor cell~\cite{cai2020,yu2022}, which is made by the anodic bonding technique, and filled with enriched $^{85}$Rb atoms and 150~Torr N$_{2}$ gases. The cavity consists of two cylindrical mirrors with a curvature of 100 mm, a diameter of 12.7 mm, and a thickness of 2.5 mm. The distance between the two mirrors is 19.3 mm, and the relative angle between their symmetrical axes is 52.2$^\circ$. This cell is placed on a three-dimensionally-printed optical platform and heated to a temperature around 85$^\circ$C by running an ac current through ceramic heaters. This magnetometer sensor sits in the middle of five-layer mu-metal shields. Solenoid coils and two sets of orthogonal cosine-theta coils~\cite{jacobsthesis} inside the innermost shield are used to control the bias field, and extra coils are added to control the field gradient.

Both the pump and probe beams for the sensor are generated from distributed-Bragg-reflector diode lasers, and fiber coupled to the sensor platform. The pump laser, resonant with Rb D1 transition, is power modulated by an acoustic-optical modulator (AOM) with a modulation frequency of $\omega$ and a duty cycle of 20\%, and then is split into two separated beams before being independently sent to the sensor. Both pump beams are circularly polarized and enter the cell from the $x$ direction, with the same beam diameter of 5~mm, and a separation between the beam centers of 1.2~cm. There is a hole with a diameter of 2.5 mm in the center of the front mirror of the multipass cell, and a linearly polarized probe beam, with a beam power of 1.6~mW and a blue detuning of 50~GHz from the D1 line, enters and exits the cavity from the same hole with 21 reflections inside the cavity. The polarization of the transmitted probe beam is analyzed by a polarization beam splitter (PBS), and a pair of differential photodiode detectors (PDs). All electrical signals pass through a printed circuit board (PCB), which is attached with the sensor. A shielded cable is used to transfer most of the electrical signals to the control electronics.

\begin{figure}[htb]
\includegraphics[width=3in]{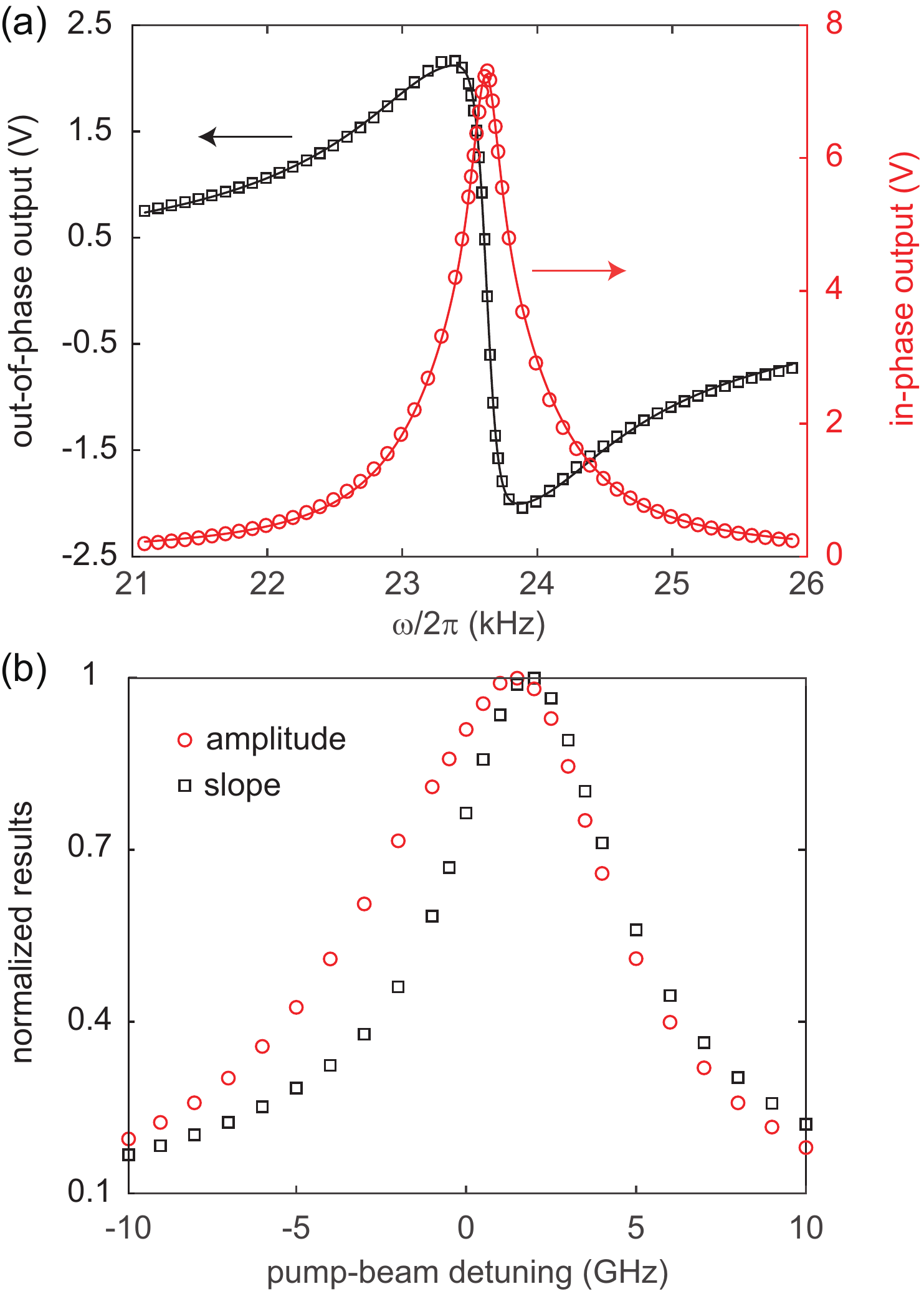}
\caption{\label{fig:scalar} Plot~(a) shows the demodulation outputs as a function of $\omega$, and the lines are fitting functions using Eq.~\eqref{eq:fit}. Plot~(b) shows the normalized amplitude ($|a_1+a_2|/|a_1+a_2|_{max}$) and slope ($|c_1+c_2|/|c_1+c_2|_{max}$) of the magnetometer signal as a function of pump beam detuning, where the resonance point is found out by fitting an absorption profile of a linearly polarized beam. In both plots, the cell temperature is 80$^\circ$C, the pump beam power is 2.25 mW, and the bias field is 5.0~$\mu$T along the $z$ direction.}
\end{figure}

When one of the pump beams is blocked, the sensor works as a scalar magnetometer. The current signals from PDs are converted to voltage signals by trans-impedance amplifiers (TIA), and are demodulated by a lock-in amplifier (LIA) as shown in Fig.~\ref{fig:setup}~(b). As the pump beam amplitude modulation frequency $\omega$ is scanned across the Larmor precession frequency $\omega_L$, the in-phase and out-of-phase outputs from LIA show a Lorentzian and a dispersion line shape, respectively (see Fig.~\ref{fig:scalar}~(a)). Due to the fact that the probe beam converges and diverges many times inside the Herriott cavity, the line shapes from the probe signal normally can not be described by a function with a single line width~\cite{li2011}. From the experience, we find that the data can be well fitted by a sum of two Lorentzian or dispersion functions~\cite{li2011,yu2022},
\begin{eqnarray}~\label{eq:fit}
f_1(\omega)&=&\sum_{m=1}^2a_m\frac{(\frac{\Gamma_m}{2})^2}{(\omega-\omega_0)^2+(\frac{\Gamma_m}{2})^2}+b_1,\nonumber\\
f_2(\omega)&=&\sum_{m=1}^2c_m\frac{(\frac{\Gamma_m}{2})^2(\omega-\omega_0)}{(\omega-\omega_0)^2+(\frac{\Gamma_m}{2})^2}+b_2.
\end{eqnarray}
Here, we can qualitatively understand that the two fitting functions correspond to dividing the beam patterns inside the cavities to two groups according to beam sizes. The fitting results of a group of in-phase and out-of-phase signals are consistent with each other. Both the amplitude of the in-phase resonant response ($f_1(\omega_0)=a_1+a_2$) and the slope of the out-of-phase resonant response ($|\partial f_2(\omega)/\partial \omega|=|c_1+c_2|$) can be used to characterize the scalar magnetometer signal.

When scanning the pump beam detuning while keeping other parameters same, we find that the largest magnetometer signal appears at a pump beam detuning around 2~GHz, as shown in Fig.~\ref{fig:scalar}~(b). Here, the atomic transition resonance frequency refers to the one pressure shifted compared with the case without buffer gases~\cite{romalis1997}, which is the same in the rest of the paper. Since the full width of the Rb D1 transition is pressure broadened to be around 4~GHz~\cite{romalis1997} due to the buffer gas, we can neglected the hyperfine splittings of Rb $5P_{1/2}$ state ($\sim 0.4$~GHz). The ground hyperfine splitting is 3.0 GHz, and the transition resonance point (zero detuning point in Fig.~\ref{fig:scalar}(b)) corresponds to the laser frequency in between the transitions $|5S_{1/2}, F=2\rangle \rightarrow |5P_{1/2}\rangle$ and $|5S_{1/2}, F=3\rangle \rightarrow |5P_{1/2}\rangle$. In cases that the pump beam is blue detuned, the laser is prone to drive the former transition; while the pump beam is red detuned, the laser is prone to drive the latter transition. Considering the redistribution of atoms due to spin-exchange collisions and quenching effects, there are more atoms populated in the ground dark state (a stretched state of $|5S_{1/2}, F=3\rangle$) when the lower hyperfine states are resonantly driven~\cite{scholtes2011}. However, due to the light shift effect induced by the off-resonant pump beams, we still choose to keep pump beam on resonance with Rb D1 transition in normal operations.

\begin{figure}[htb]
\includegraphics[width=3in]{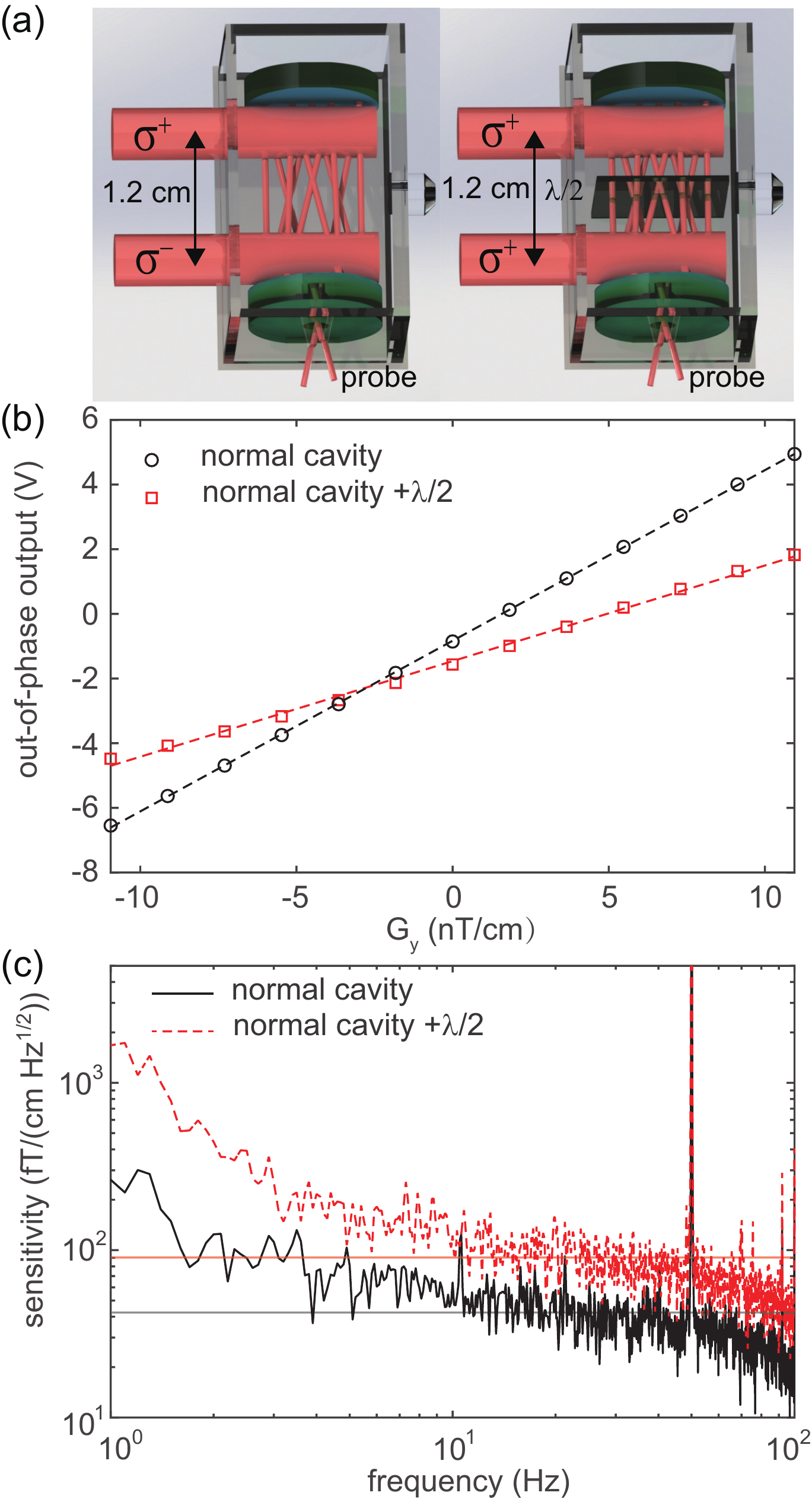}
\caption{\label{fig:result1} Plot~(a) shows two configurations for the atomic gradiometer. Plot~(b) and (c) show the gradiometer response to $G_y$ and the gradiometer sensitivity, respectively. The data for the normal cavity are taken with a bias field of 5.0~$\mu$T along the $z$ axis, each pump beam power around 5.5 mW, and a cell temperature of 85 $^\circ$C. The data for the cavity with a half-wave plate are taken with the same parameters except that the cell temperature is 80$^\circ$C. The dashed lines in plot (b) are linear fitting results, and the two horizontal lines in plot (c) denote the average noise level in the range of 15~Hz to 40~Hz.}
\end{figure}

When the other pump beam is unblocked, there are two isolated Bell-Bloom scalar magnetometers inside the same cell, considering the fact that the diffusion distance of atoms within the atomic depolarization time is much less than the separation between the two magnetometers. Since both magnetometers share the same probe beam, the signals of the magnetometers add up if the pump beam polarizations are the same, and the signals are subtracted if the pump polarizations are opposite. The latter case, which is shown in the left part of Fig.~\ref{fig:result1}(a), corresponds to a magnetic gradiometer. Since this gradiometer is built on the scalar magnetometers, it can only detect the total field gradients.

Moreover, the separation between the pump beams is along the $y$ direction, and the sensor using the demodulation processes in Fig.~\ref{fig:setup}(b) can only measure $G_y$ . This conclusion is true for all bias field directions, and in this section, we demonstrate the measurements of $G_y$ by setting the bias field along the $z$ axis for convenience. As indicated by Eqs.~\eqref{eq:Py} and ~\eqref{eq:Py2}, we can directly read the gradiometer signal from the out-of-phase output, which shows a linear dependence on $G_y$ (See Fig.~\ref{fig:result1}(b)). While the powers of the two beams should be identical in the ideal case, they are different within 10\% in practice due to slight difference in various experimental conditions for the two beams, such as the difference in beam sizes and the transmission conditions of the cell window. In the experiment, we fixed the power of one pump beam, tuned the power of the other beam, and optimized the subtraction result from the gradiometer~\cite{lucivero2021}. The gradient field sensitivity on $G_y$ is measured to be 42 fT/cm/Hz$^{1/2}$ over the frequency range of 15~Hz to 40~Hz, as shown in Fig.~\ref{fig:result1}(c). In this work, the sensitivity results are calculated from the power spectral density of the data using the Hanning window.

From the discussions in Sec.~\ref{sec:hfplate}, we can realize the magnetic gradiometer with a different configuration. As shown in the right part of Fig.~\ref{fig:result1}(a), both pump beams have the same polarization while a true zero-order half-wave plate is added in the center of the multipass cell~\cite{yu2022}, which helps to realize a differential detection of the two polarized atomic ensembles. This configuration has an advantage over the previous one that it can eliminate common-mode noise from the light shift, if the pump beam frequency is detuned from the resonance. In practice, the gradient field sensitivity of $G_y$ in this configuration is measured to be 90 fT/cm/Hz$^{1/2}$ over the frequency range of 15~Hz to 40~Hz. The fact that the latter configuration shows worse sensitivity results are probably due to two facts, one is that the working temperature of the half-wave plate limits the maximum cell temperature~\cite{yu2022}, and the other one is that the quality of the half-wave plate degrades in the heated environment, and in turn the quality of the beam polarization degrades as the number of passes of the beam through the plate (22 times in maximum) increases.

\section{Partial measurement of the field gradient tensor}

\begin{figure}[htb]
\includegraphics[width=3in]{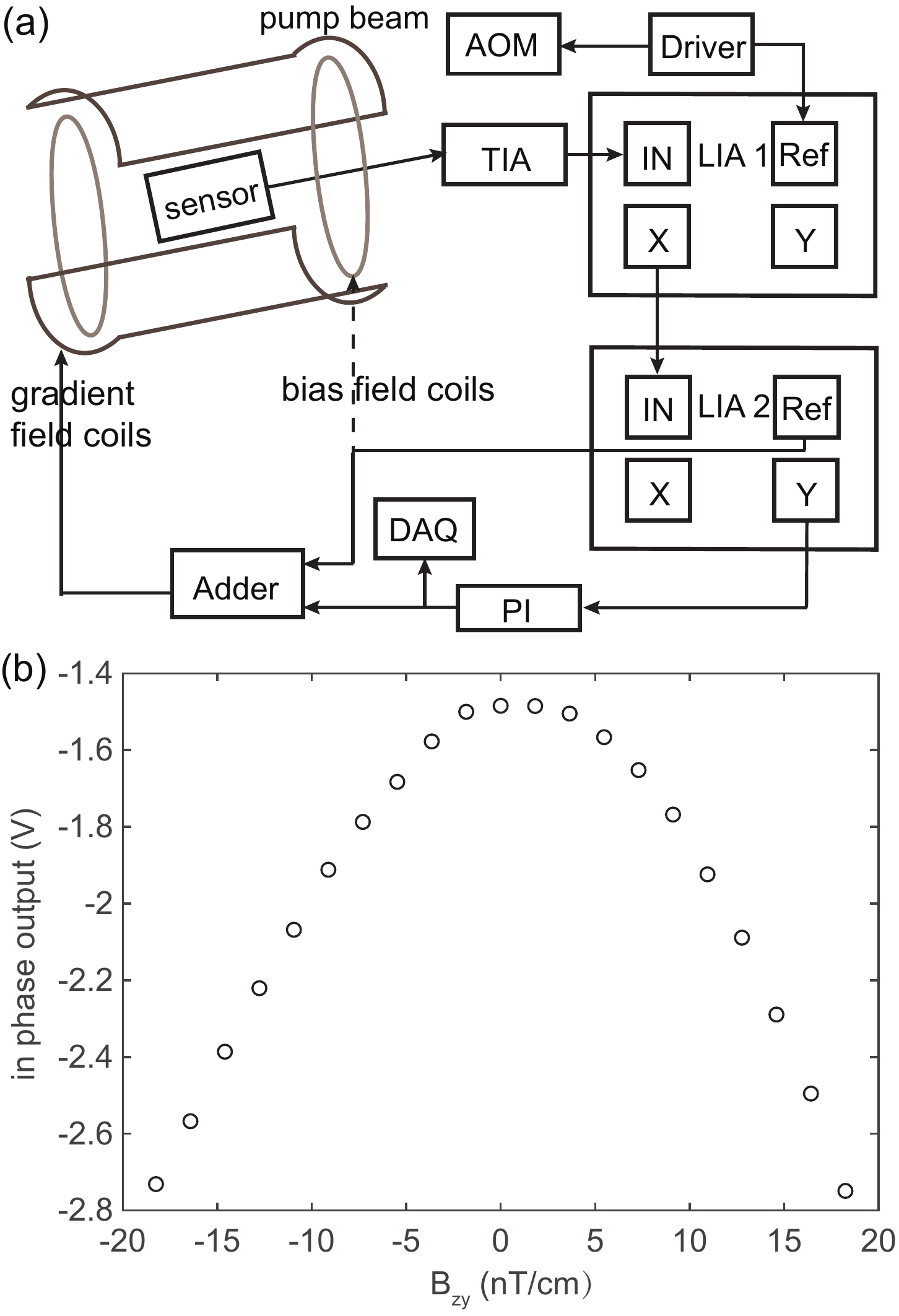}
\caption{\label{fig:flow} Plot~(a) shows signal processing of measuring elements of the field gradient tensor, where the sensor in the center of the coil system corresponds to the atomic gradiometer with the same configuration as the  left one in Fig.~\ref{fig:result1}(a). Plot~(b) shows the out-of-phase output of LIA1 as a function of $B_{zy}$, with a bias field of $\boldsymbol{B}=(\hat{x}+\hat{y}+\hat{z})5.8$~$\mu$T, a cell temperature of 85~$^\circ$C, and each pump beam power around 5.5~mW.}
\end{figure}

To measure the field gradient tensor, the sensor needs to have the ability to distinguish different vector components that contribute to the total field magnitude. One convenient way to reach this goal is to add modulation fields at different directions~\cite{patton2014,zheng2020,wang2022}, and use the frequencies or phases of the modulation fields to selectively measure and control the field gradient tensor elements. In this way, while keeping components in Fig.~\ref{fig:setup}(b) as the hardware for first part of the data processes, we need extra LIAs for successive demodulations as shown in Fig.~\ref{fig:flow}(a). We choose the pump beam configuration on the left of Fig.~\ref{fig:result1}(a) for this application.

While the pump beam powers are modulated at the Larmor frequencies, according to Eq.~\eqref{eq:Py2}, adding a field gradient modifies $\omega_L$, and in turn affects the $\omega-\omega_L$ term in the denominator of the equation. However, the value of the in-phase output is independent of the sign of the additional field gradient. Therefore, as shown in Fig.~\ref{fig:flow}(b), the in-phase output of the first LIA (LIA1 in Fig.~\ref{fig:flow}(a)) is a symmetric function of $B_{zy}$ with a bias field of $\boldsymbol{B}=(\hat{x}+\hat{y}+\hat{z})5.8$~$\mu$T. This result is similar to the case of a magnetometer based on a single circularly polarized beam~\cite{sheng17}, where the transmission power of the beam is a symmetric function of the transverse field. In the analogy, we can adopt the closed-loop operation scheme in Ref.~\cite{sheng17} to measure the field gradient tensor elements. For example, we apply a modulation field with a frequency of $\omega_d$ to the $B_{zy}$ coils. The signal that is simultaneously modulated at $\omega$ and $\omega_d$ are extracted by successive demodulations, whose final out is proportional to the bias value of $B_{zy}$. By passing this output to a proportional-integral (PI) controller with a setting point of zero and feeding back on the $B_{zy}$ coils, we can constantly null the bias value of $B_{zy}$ in the place of the sensor. The feedback value from the PI output is converted to the value of $B_{zy}$ using the current-to-field calibration factor of the gradient coil, which is measured by a
fluxgate, and recorded as the gradiometer output.

\begin{figure}[htb]
\includegraphics[width=3in]{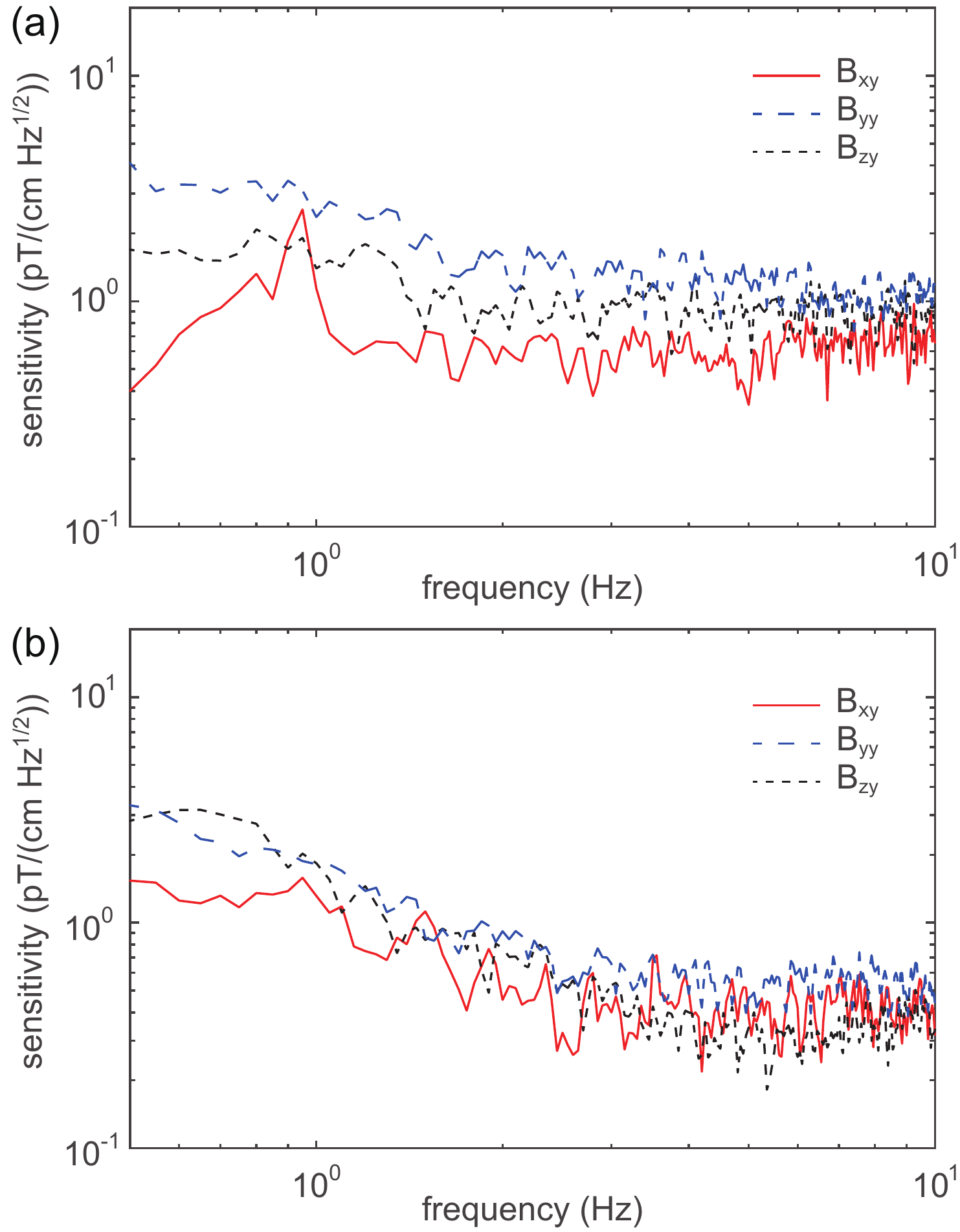}
\caption{\label{fig:result2}  Plot~(a) shows the gradient field sensitivities for $B_{xy}$, $B_{yy}$, and $B_{zy}$, where the experimental conditions are the same as those in Fig.~\ref{fig:flow}(b). Plot~(b) shows the same results as plot~(a), except that the modulation fields are applied via the bias field coils.}
\end{figure}

Using the scheme described above, we apply modulation fields on $B_{xy}$, $B_{yy}$ and $B_{zy}$ coils, all of which share similar amplitudes around 10~nT/cm, but have different modulation frequencies, ranging from 175~Hz to 297~Hz. To calibrate the cross-talk between either two of the above three channels that measure the field gradient tensor elements, we operate both channels in the closed-loop mode, scan the current in the gradient field coil for one channel, and record the response of the other channel. In this way, the cross talks between channels are calibrated to be less than 0.3\%. As shown in Fig.~\ref{fig:result2}~(a), the closed-loop gradient field sensitivities for these three elements are 0.67~pT/cm/Hz$^{1/2}$, 0.89~pT/cm/Hz$^{1/2}$, and 0.87~pT/cm/Hz$^{1/2}$ over the frequency range of 5~Hz to 10~Hz.

Another way to realize the same functions is to apply the modulation fields along the bias field coils, instead of the gradient field coils, while keeping the other parts unchanged. Compared with the former method, this configuration avoids gradient field modulations which reduce the atomic depolarization time. The experimental parameters are the same as in the aforementioned method, except that amplitudes of the modulation fields applied on the three-axis bias-field coils are around 50~nT. The cross talks between channels in this case are calibrated to be less than 0.2\%, and the closed-loop gradient field sensitivities for $B_{xy}$, $B_{yy}$ and $B_{zy}$ are improved to 0.35~pT/cm/Hz$^{1/2}$, 0.49~pT/cm/Hz$^{1/2}$, and 0.34~pT/cm/Hz$^{1/2}$ over the frequency range of 5~Hz to 10~Hz, as shown in Fig.~\ref{fig:result2}(b).

\section{Conclusion}
In summary, we have developed a versatile Herriott-cavity-assisted atomic gradiometer, using two Bell-Bloom optical pumping beams and a single probe beam. We demonstrated its applications in measuring some elements of the total field gradient and the field gradient tensor.

Currently, the sensor sensitivities on the field gradient components are mainly limited by the noises from the pump beams, which can be eliminated using a pulsed pump-probe scheme~\cite{limes2020,lucivero2021,lee2021}. Due to the two pump beams for the sensor are placed in parallel along the $y$ axis in this paper, the sensor can only measure the field gradient elements related to the $y$ axis. For a full measurement of the total field gradient and the gradient tensor, we can add more pairs of pump beams in parallel along the $x$ and $z$ directions for differential detections of atomic ensembles in these two axes, or adding more gradiometers with different orientations so that the sensitive directions of such gradiometers can cover all three axes.

\section*{Acknowledgements}
This work was partially carried out at the USTC Center for Micro and Nanoscale Research and Fabrication. This work was supported by National Natural Science Foundation of China (Grant No. 11974329), and Scientific Instrument and Equipment Development Projects, CAS (NO. YJKYYQ20200043).


\begin{thebibliography}{46}%
\makeatletter
\providecommand \@ifxundefined [1]{%
 \@ifx{#1\undefined}
}%
\providecommand \@ifnum [1]{%
 \ifnum #1\expandafter \@firstoftwo
 \else \expandafter \@secondoftwo
 \fi
}%
\providecommand \@ifx [1]{%
 \ifx #1\expandafter \@firstoftwo
 \else \expandafter \@secondoftwo
 \fi
}%
\providecommand \natexlab [1]{#1}%
\providecommand \enquote  [1]{``#1''}%
\providecommand \bibnamefont  [1]{#1}%
\providecommand \bibfnamefont [1]{#1}%
\providecommand \citenamefont [1]{#1}%
\providecommand \href@noop [0]{\@secondoftwo}%
\providecommand \href [0]{\begingroup \@sanitize@url \@href}%
\providecommand \@href[1]{\@@startlink{#1}\@@href}%
\providecommand \@@href[1]{\endgroup#1\@@endlink}%
\providecommand \@sanitize@url [0]{\catcode `\\12\catcode `\$12\catcode
  `\&12\catcode `\#12\catcode `\^12\catcode `\_12\catcode `\%12\relax}%
\providecommand \@@startlink[1]{}%
\providecommand \@@endlink[0]{}%
\providecommand \url  [0]{\begingroup\@sanitize@url \@url }%
\providecommand \@url [1]{\endgroup\@href {#1}{\urlprefix }}%
\providecommand \urlprefix  [0]{URL }%
\providecommand \Eprint [0]{\href }%
\providecommand \doibase [0]{https://doi.org/}%
\providecommand \selectlanguage [0]{\@gobble}%
\providecommand \bibinfo  [0]{\@secondoftwo}%
\providecommand \bibfield  [0]{\@secondoftwo}%
\providecommand \translation [1]{[#1]}%
\providecommand \BibitemOpen [0]{}%
\providecommand \bibitemStop [0]{}%
\providecommand \bibitemNoStop [0]{.\EOS\space}%
\providecommand \EOS [0]{\spacefactor3000\relax}%
\providecommand \BibitemShut  [1]{\csname bibitem#1\endcsname}%
\let\auto@bib@innerbib\@empty
%</preamble>
\bibitem [{\citenamefont {Schmidt}\ and\ \citenamefont
  {Clark}(2006)}]{schmidt2006}%
  \BibitemOpen
  \bibfield  {author} {\bibinfo {author} {\bibfnamefont {P.}~\bibnamefont
  {Schmidt}}\ and\ \bibinfo {author} {\bibfnamefont {D.}~\bibnamefont
  {Clark}},\ }\bibfield  {title} {\bibinfo {title} {{The magnetic gradient
  tensor : Its properties and uses in source characterization}},\ }\href
  {https://doi.org/10.1190/1.2164759} {\bibfield  {journal} {\bibinfo
  {journal} {The Leading Edge}\ }\textbf {\bibinfo {volume} {25}},\ \bibinfo
  {pages} {75} (\bibinfo {year} {2006})}\BibitemShut {NoStop}%
\bibitem [{\citenamefont {{Getmag Team}}(2004)}]{getmag2004}%
  \BibitemOpen
  \bibfield  {author} {\bibinfo {author} {\bibnamefont {{Getmag Team}}},\
  }\bibfield  {title} {\bibinfo {title} {{GETMAG} – a new magnetic tensor
  gradiometer for exploration},\ }\href {https://doi.org/10.1071/ASEG2004ab047}
  {\bibfield  {journal} {\bibinfo  {journal} {ASEG Extended Abstracts}\
  }\textbf {\bibinfo {volume} {2004}},\ \bibinfo {pages} {1} (\bibinfo {year}
  {2004})}\BibitemShut {NoStop}%
\bibitem [{\citenamefont {Hirota}\ \emph {et~al.}(1997)\citenamefont {Hirota},
  \citenamefont {Nanaura}, \citenamefont {Teranishi},\ and\ \citenamefont
  {Kishigami}}]{hirota1997}%
  \BibitemOpen
  \bibfield  {author} {\bibinfo {author} {\bibfnamefont {M.}~\bibnamefont
  {Hirota}}, \bibinfo {author} {\bibfnamefont {K.}~\bibnamefont {Nanaura}},
  \bibinfo {author} {\bibfnamefont {Y.}~\bibnamefont {Teranishi}},\ and\
  \bibinfo {author} {\bibfnamefont {T.}~\bibnamefont {Kishigami}},\ }\bibfield
  {title} {\bibinfo {title} {{SQUID} gradiometers for a fundamental study of
  underwater magnetic detection},\ }\href {https://doi.org/10.1109/77.621705}
  {\bibfield  {journal} {\bibinfo  {journal} {IEEE Transactions on Applied
  Superconductivity}\ }\textbf {\bibinfo {volume} {7}},\ \bibinfo {pages}
  {2327} (\bibinfo {year} {1997})}\BibitemShut {NoStop}%
\bibitem [{\citenamefont {Wickerham}(1954)}]{wickerham1954}%
  \BibitemOpen
  \bibfield  {author} {\bibinfo {author} {\bibfnamefont {W.~E.}\ \bibnamefont
  {Wickerham}},\ }\bibfield  {title} {\bibinfo {title} {The gulf airborne
  magnetic gradiometer},\ }\href {https://doi.org/10.1190/1.1437955} {\bibfield
   {journal} {\bibinfo  {journal} {Geophysics}\ }\textbf {\bibinfo {volume}
  {19}},\ \bibinfo {pages} {116} (\bibinfo {year} {1954})}\BibitemShut
  {NoStop}%
\bibitem [{\citenamefont {Overton}(1981)}]{overton1981}%
  \BibitemOpen
  \bibfield  {author} {\bibinfo {author} {\bibfnamefont {J.}~\bibnamefont
  {Overton}, \bibfnamefont {W.~C.}},\ }\bibfield  {title} {\bibinfo {title}
  {Detection of a thin sheet magnetic anomaly by {SQUID} gradiometer systems:
  Possibility of hydrofracture azimuth determination *},\ }in\ \href
  {https://doi.org/10.1190/1.9781560802518} {\emph {\bibinfo {booktitle} {SQUID
  Applications to Geophysics}}}\ (\bibinfo  {publisher} {Society of Exploration
  Geophysicists},\ \bibinfo {year} {1981})\BibitemShut {NoStop}%
\bibitem [{\citenamefont {Clark}(2012)}]{clark2012}%
  \BibitemOpen
  \bibfield  {author} {\bibinfo {author} {\bibfnamefont {D.~A.}\ \bibnamefont
  {Clark}},\ }\bibfield  {title} {\bibinfo {title} {New methods for
  interpretation of magnetic gradient tensor data},\ }\href
  {https://doi.org/10.1071/ASEG2012ab081} {\bibfield  {journal} {\bibinfo
  {journal} {ASEG Extended Abstracts}\ }\textbf {\bibinfo {volume} {2012}},\
  \bibinfo {pages} {1} (\bibinfo {year} {2012})}\BibitemShut {NoStop}%
\bibitem [{\citenamefont {Acuna}(2002)}]{acuna2002}%
  \BibitemOpen
  \bibfield  {author} {\bibinfo {author} {\bibfnamefont {M.~H.}\ \bibnamefont
  {Acuna}},\ }\bibfield  {title} {\bibinfo {title} {Space-based
  magnetometers},\ }\href {https://doi.org/10.1063/1.1510570} {\bibfield
  {journal} {\bibinfo  {journal} {Rev. Sci. Instr.}\ }\textbf {\bibinfo
  {volume} {73}},\ \bibinfo {pages} {3717} (\bibinfo {year}
  {2002})}\BibitemShut {NoStop}%
\bibitem [{\citenamefont {Canciani}\ and\ \citenamefont
  {Raquet}(2016)}]{canciani2016}%
  \BibitemOpen
  \bibfield  {author} {\bibinfo {author} {\bibfnamefont {A.}~\bibnamefont
  {Canciani}}\ and\ \bibinfo {author} {\bibfnamefont {J.}~\bibnamefont
  {Raquet}},\ }\bibfield  {title} {\bibinfo {title} {Absolute positioning using
  the earth's magnetic anomaly field},\ }\href
  {https://doi.org/https://doi.org/10.1002/navi.138op} {\bibfield  {journal}
  {\bibinfo  {journal} {Journal of The Institute of Navigation}\ }\textbf
  {\bibinfo {volume} {63}},\ \bibinfo {pages} {111} (\bibinfo {year}
  {2016})}\BibitemShut {NoStop}%
\bibitem [{\citenamefont {Zhang}\ \emph {et~al.}(2020)\citenamefont {Zhang},
  \citenamefont {Xiao}, \citenamefont {Ding}, \citenamefont {Feng},
  \citenamefont {Peng}, \citenamefont {Shen}, \citenamefont {Sun},
  \citenamefont {Wu}, \citenamefont {Wu}, \citenamefont {Yang}, \citenamefont
  {Zheng}, \citenamefont {Zhang}, \citenamefont {Chen},\ and\ \citenamefont
  {Guo}}]{zhangrui2020}%
  \BibitemOpen
  \bibfield  {author} {\bibinfo {author} {\bibfnamefont {R.}~\bibnamefont
  {Zhang}}, \bibinfo {author} {\bibfnamefont {W.}~\bibnamefont {Xiao}},
  \bibinfo {author} {\bibfnamefont {Y.}~\bibnamefont {Ding}}, \bibinfo {author}
  {\bibfnamefont {Y.}~\bibnamefont {Feng}}, \bibinfo {author} {\bibfnamefont
  {X.}~\bibnamefont {Peng}}, \bibinfo {author} {\bibfnamefont {L.}~\bibnamefont
  {Shen}}, \bibinfo {author} {\bibfnamefont {C.}~\bibnamefont {Sun}}, \bibinfo
  {author} {\bibfnamefont {T.}~\bibnamefont {Wu}}, \bibinfo {author}
  {\bibfnamefont {Y.}~\bibnamefont {Wu}}, \bibinfo {author} {\bibfnamefont
  {Y.}~\bibnamefont {Yang}}, \bibinfo {author} {\bibfnamefont {Z.}~\bibnamefont
  {Zheng}}, \bibinfo {author} {\bibfnamefont {X.}~\bibnamefont {Zhang}},
  \bibinfo {author} {\bibfnamefont {J.}~\bibnamefont {Chen}},\ and\ \bibinfo
  {author} {\bibfnamefont {H.}~\bibnamefont {Guo}},\ }\bibfield  {title}
  {\bibinfo {title} {Recording brain activities in unshielded earth's field
  with optically pumped atomic magnetometers},\ }\href
  {https://doi.org/10.1126/sciadv.aba8792} {\bibfield  {journal} {\bibinfo
  {journal} {Science Advances}\ }\textbf {\bibinfo {volume} {6}},\ \bibinfo
  {pages} {eaba8792} (\bibinfo {year} {2020})}\BibitemShut {NoStop}%
\bibitem [{\citenamefont {Limes}\ \emph {et~al.}(2020)\citenamefont {Limes},
  \citenamefont {Foley}, \citenamefont {Kornack}, \citenamefont {Caliga},
  \citenamefont {McBride}, \citenamefont {Braun}, \citenamefont {Lee},
  \citenamefont {Lucivero},\ and\ \citenamefont {Romalis}}]{limes2020}%
  \BibitemOpen
  \bibfield  {author} {\bibinfo {author} {\bibfnamefont {M.}~\bibnamefont
  {Limes}}, \bibinfo {author} {\bibfnamefont {E.}~\bibnamefont {Foley}},
  \bibinfo {author} {\bibfnamefont {T.}~\bibnamefont {Kornack}}, \bibinfo
  {author} {\bibfnamefont {S.}~\bibnamefont {Caliga}}, \bibinfo {author}
  {\bibfnamefont {S.}~\bibnamefont {McBride}}, \bibinfo {author} {\bibfnamefont
  {A.}~\bibnamefont {Braun}}, \bibinfo {author} {\bibfnamefont
  {W.}~\bibnamefont {Lee}}, \bibinfo {author} {\bibfnamefont {V.}~\bibnamefont
  {Lucivero}},\ and\ \bibinfo {author} {\bibfnamefont {M.}~\bibnamefont
  {Romalis}},\ }\bibfield  {title} {\bibinfo {title} {Portable magnetometry for
  detection of biomagnetism in ambient environments},\ }\href
  {https://doi.org/10.1103/PhysRevApplied.14.011002} {\bibfield  {journal}
  {\bibinfo  {journal} {Phys. Rev. Applied}\ }\textbf {\bibinfo {volume}
  {14}},\ \bibinfo {pages} {011002} (\bibinfo {year} {2020})}\BibitemShut
  {NoStop}%
\bibitem [{\citenamefont {Stolz}\ \emph {et~al.}(2006)\citenamefont {Stolz},
  \citenamefont {Zakosarenko}, \citenamefont {Schulz}, \citenamefont {Chwala},
  \citenamefont {Fritzsch}, \citenamefont {Meyer},\ and\ \citenamefont
  {K\''{o}stlin}}]{stolz2006}%
  \BibitemOpen
  \bibfield  {author} {\bibinfo {author} {\bibfnamefont {R.}~\bibnamefont
  {Stolz}}, \bibinfo {author} {\bibfnamefont {V.}~\bibnamefont {Zakosarenko}},
  \bibinfo {author} {\bibfnamefont {M.}~\bibnamefont {Schulz}}, \bibinfo
  {author} {\bibfnamefont {A.}~\bibnamefont {Chwala}}, \bibinfo {author}
  {\bibfnamefont {L.}~\bibnamefont {Fritzsch}}, \bibinfo {author}
  {\bibfnamefont {H.-G.}\ \bibnamefont {Meyer}},\ and\ \bibinfo {author}
  {\bibfnamefont {E.}~\bibnamefont {K\''{o}stlin}},\ }\bibfield  {title}
  {\bibinfo {title} {{Magnetic full-tensor SQUID gradiometer system for
  geophysical applications}},\ }\href {https://doi.org/10.1190/1.2172308}
  {\bibfield  {journal} {\bibinfo  {journal} {The Leading Edge}\ }\textbf
  {\bibinfo {volume} {25}},\ \bibinfo {pages} {178} (\bibinfo {year}
  {2006})}\BibitemShut {NoStop}%
\bibitem [{\citenamefont {Wynn}\ \emph {et~al.}(1975)\citenamefont {Wynn},
  \citenamefont {Frahm}, \citenamefont {Carroll}, \citenamefont {Clark},
  \citenamefont {Wellhoner},\ and\ \citenamefont {Wynn}}]{wynn1975}%
  \BibitemOpen
  \bibfield  {author} {\bibinfo {author} {\bibfnamefont {W.}~\bibnamefont
  {Wynn}}, \bibinfo {author} {\bibfnamefont {C.}~\bibnamefont {Frahm}},
  \bibinfo {author} {\bibfnamefont {P.}~\bibnamefont {Carroll}}, \bibinfo
  {author} {\bibfnamefont {R.}~\bibnamefont {Clark}}, \bibinfo {author}
  {\bibfnamefont {J.}~\bibnamefont {Wellhoner}},\ and\ \bibinfo {author}
  {\bibfnamefont {M.}~\bibnamefont {Wynn}},\ }\bibfield  {title} {\bibinfo
  {title} {Advanced superconducting gradiometer/magnetometer arrays and a novel
  signal processing technique},\ }\href
  {https://doi.org/10.1109/TMAG.1975.1058672} {\bibfield  {journal} {\bibinfo
  {journal} {IEEE Transactions on Magnetics}\ }\textbf {\bibinfo {volume}
  {11}},\ \bibinfo {pages} {701} (\bibinfo {year} {1975})}\BibitemShut
  {NoStop}%
\bibitem [{\citenamefont {Clem}\ \emph {et~al.}(1996)\citenamefont {Clem},
  \citenamefont {Kekelis}, \citenamefont {Lathrop}, \citenamefont {Overway},\
  and\ \citenamefont {Wynn}}]{clem1996}%
  \BibitemOpen
  \bibfield  {author} {\bibinfo {author} {\bibfnamefont {T.~R.}\ \bibnamefont
  {Clem}}, \bibinfo {author} {\bibfnamefont {G.~J.}\ \bibnamefont {Kekelis}},
  \bibinfo {author} {\bibfnamefont {J.~D.}\ \bibnamefont {Lathrop}}, \bibinfo
  {author} {\bibfnamefont {D.~J.}\ \bibnamefont {Overway}},\ and\ \bibinfo
  {author} {\bibfnamefont {W.~M.}\ \bibnamefont {Wynn}},\ }\bibfield  {title}
  {\bibinfo {title} {Superconducting magnetic gradiometers for mobile
  applications with an emphasis on ordnance detection},\ }in\ \href@noop {}
  {\emph {\bibinfo {booktitle} {SQUID Sensors: Fundamentals, Fabrication and
  Applications}}},\ \bibinfo {editor} {edited by\ \bibinfo {editor}
  {\bibfnamefont {H.}~\bibnamefont {Weinstock}}}\ (\bibinfo  {publisher}
  {Springer Dordrecht},\ \bibinfo {year} {1996})\ pp.\ \bibinfo {pages}
  {517--568}\BibitemShut {NoStop}%
\bibitem [{\citenamefont {Keenan}\ \emph {et~al.}(2022)\citenamefont {Keenan},
  \citenamefont {Clark},\ and\ \citenamefont {Leslie}}]{keenan2022}%
  \BibitemOpen
  \bibfield  {author} {\bibinfo {author} {\bibfnamefont {S.~T.}\ \bibnamefont
  {Keenan}}, \bibinfo {author} {\bibfnamefont {D.~A.}\ \bibnamefont {Clark}},\
  and\ \bibinfo {author} {\bibfnamefont {K.~E.}\ \bibnamefont {Leslie}},\
  }\bibfield  {title} {\bibinfo {title} {Method for full magnetic gradient
  tensor detection from a single {HTS} gradiometer},\ }\href
  {https://doi.org/10.1088/1361-6668/ac5016} {\bibfield  {journal} {\bibinfo
  {journal} {Superconductor Science and Technology}\ }\textbf {\bibinfo
  {volume} {35}},\ \bibinfo {pages} {045005} (\bibinfo {year}
  {2022})}\BibitemShut {NoStop}%
\bibitem [{\citenamefont {Wiegert}\ \emph {et~al.}(2007)\citenamefont
  {Wiegert}, \citenamefont {Oeschger},\ and\ \citenamefont
  {Tuovila}}]{wiegert2007}%
  \BibitemOpen
  \bibfield  {author} {\bibinfo {author} {\bibfnamefont {R.}~\bibnamefont
  {Wiegert}}, \bibinfo {author} {\bibfnamefont {J.}~\bibnamefont {Oeschger}},\
  and\ \bibinfo {author} {\bibfnamefont {E.}~\bibnamefont {Tuovila}},\
  }\bibfield  {title} {\bibinfo {title} {Demonstration of a novel man-portable
  magnetic star technology for real time localization of unexploded ordnance},\
  }in\ \href {https://doi.org/10.1109/OCEANS.2007.4449229} {\emph {\bibinfo
  {booktitle} {OCEANS 2007}}}\ (\bibinfo {year} {2007})\ pp.\ \bibinfo {pages}
  {1--7}\BibitemShut {NoStop}%
\bibitem [{\citenamefont {Pang}\ \emph {et~al.}(2014)\citenamefont {Pang},
  \citenamefont {Pan}, \citenamefont {Wan}, \citenamefont {Chen}, \citenamefont
  {Zhu},\ and\ \citenamefont {Luo}}]{pang2014}%
  \BibitemOpen
  \bibfield  {author} {\bibinfo {author} {\bibfnamefont {H.}~\bibnamefont
  {Pang}}, \bibinfo {author} {\bibfnamefont {M.}~\bibnamefont {Pan}}, \bibinfo
  {author} {\bibfnamefont {C.}~\bibnamefont {Wan}}, \bibinfo {author}
  {\bibfnamefont {J.}~\bibnamefont {Chen}}, \bibinfo {author} {\bibfnamefont
  {X.}~\bibnamefont {Zhu}},\ and\ \bibinfo {author} {\bibfnamefont
  {F.}~\bibnamefont {Luo}},\ }\bibfield  {title} {\bibinfo {title} {Integrated
  compensation of magnetometer array magnetic distortion field and improvement
  of magnetic object localization},\ }\href
  {https://doi.org/10.1109/TGRS.2013.2291839} {\bibfield  {journal} {\bibinfo
  {journal} {IEEE Transactions on Geoscience and Remote Sensing}\ }\textbf
  {\bibinfo {volume} {52}},\ \bibinfo {pages} {5670} (\bibinfo {year}
  {2014})}\BibitemShut {NoStop}%
\bibitem [{\citenamefont {Yin}\ and\ \citenamefont {Zhang}(2018)}]{yin2018}%
  \BibitemOpen
  \bibfield  {author} {\bibinfo {author} {\bibfnamefont {G.}~\bibnamefont
  {Yin}}\ and\ \bibinfo {author} {\bibfnamefont {L.}~\bibnamefont {Zhang}},\
  }\bibfield  {title} {\bibinfo {title} {Magnetic heading compensation method
  based on magnetic interferential signal inversion},\ }\href
  {https://doi.org/https://doi.org/10.1016/j.sna.2018.03.043} {\bibfield
  {journal} {\bibinfo  {journal} {Sensors and Actuators A: Physical}\ }\textbf
  {\bibinfo {volume} {275}},\ \bibinfo {pages} {1} (\bibinfo {year}
  {2018})}\BibitemShut {NoStop}%
\bibitem [{\citenamefont {Stolz}\ \emph {et~al.}(2021)\citenamefont {Stolz},
  \citenamefont {Schmelz}, \citenamefont {Zakosarenko}, \citenamefont {Foley},
  \citenamefont {Tanabe}, \citenamefont {Xie},\ and\ \citenamefont
  {Fagaly}}]{stolz2021}%
  \BibitemOpen
  \bibfield  {author} {\bibinfo {author} {\bibfnamefont {R.}~\bibnamefont
  {Stolz}}, \bibinfo {author} {\bibfnamefont {M.}~\bibnamefont {Schmelz}},
  \bibinfo {author} {\bibfnamefont {V.}~\bibnamefont {Zakosarenko}}, \bibinfo
  {author} {\bibfnamefont {C.}~\bibnamefont {Foley}}, \bibinfo {author}
  {\bibfnamefont {K.}~\bibnamefont {Tanabe}}, \bibinfo {author} {\bibfnamefont
  {X.}~\bibnamefont {Xie}},\ and\ \bibinfo {author} {\bibfnamefont {R.~L.}\
  \bibnamefont {Fagaly}},\ }\bibfield  {title} {\bibinfo {title}
  {Superconducting sensors and methods in geophysical applications},\ }\href
  {https://doi.org/10.1088/1361-6668/abd7ce} {\bibfield  {journal} {\bibinfo
  {journal} {Superconductor Science and Technology}\ }\textbf {\bibinfo
  {volume} {34}},\ \bibinfo {pages} {033001} (\bibinfo {year}
  {2021})}\BibitemShut {NoStop}%
\bibitem [{\citenamefont {Stolz}\ \emph {et~al.}(2022)\citenamefont {Stolz},
  \citenamefont {Schiffler}, \citenamefont {Becken}, \citenamefont {Thiede},
  \citenamefont {Schneider}, \citenamefont {Chubak}, \citenamefont {Marsden},
  \citenamefont {Bergshjorth}, \citenamefont {Schaefer},\ and\ \citenamefont
  {Terblanche}}]{stolz2022}%
  \BibitemOpen
  \bibfield  {author} {\bibinfo {author} {\bibfnamefont {R.}~\bibnamefont
  {Stolz}}, \bibinfo {author} {\bibfnamefont {M.}~\bibnamefont {Schiffler}},
  \bibinfo {author} {\bibfnamefont {M.}~\bibnamefont {Becken}}, \bibinfo
  {author} {\bibfnamefont {A.}~\bibnamefont {Thiede}}, \bibinfo {author}
  {\bibfnamefont {M.}~\bibnamefont {Schneider}}, \bibinfo {author}
  {\bibfnamefont {G.}~\bibnamefont {Chubak}}, \bibinfo {author} {\bibfnamefont
  {P.}~\bibnamefont {Marsden}}, \bibinfo {author} {\bibfnamefont {A.~B.}\
  \bibnamefont {Bergshjorth}}, \bibinfo {author} {\bibfnamefont
  {M.}~\bibnamefont {Schaefer}},\ and\ \bibinfo {author} {\bibfnamefont
  {O.}~\bibnamefont {Terblanche}},\ }\bibfield  {title} {\bibinfo {title}
  {{SQUID}s for magnetic and electromagnetic methods in mineral exploration},\
  }\href {https://doi.org/10.1007/s13563-022-00333-3} {\bibfield  {journal}
  {\bibinfo  {journal} {Mineral Economics}\ }\textbf {\bibinfo {volume} {35}},\
  \bibinfo {pages} {467 } (\bibinfo {year} {2022})}\BibitemShut {NoStop}%
\bibitem [{\citenamefont {Budker}\ and\ \citenamefont {{Jackson
  Kimball}}(2013)}]{budker2013}%
  \BibitemOpen
  \bibinfo {editor} {\bibfnamefont {D.}~\bibnamefont {Budker}}\ and\ \bibinfo
  {editor} {\bibfnamefont {D.~F.}\ \bibnamefont {{Jackson Kimball}}},\ eds.,\
  \href {https://doi.org/10.1017/CBO9780511846380} {\emph {\bibinfo {title}
  {Optical Magnetometry}}}\ (\bibinfo  {publisher} {Cambridge University
  Press},\ \bibinfo {address} {Cambridge},\ \bibinfo {year} {2013})\BibitemShut
  {NoStop}%
\bibitem [{\citenamefont {Kitching}(2018)}]{kitching2018}%
  \BibitemOpen
  \bibfield  {author} {\bibinfo {author} {\bibfnamefont {J.}~\bibnamefont
  {Kitching}},\ }\bibfield  {title} {\bibinfo {title} {Chip-scale atomic
  devices},\ }\href {https://doi.org/10.1063/1.5026238} {\bibfield  {journal}
  {\bibinfo  {journal} {Appl. Phys. Rev.}\ }\textbf {\bibinfo {volume} {5}},\
  \bibinfo {pages} {031302} (\bibinfo {year} {2018})}\BibitemShut {NoStop}%
\bibitem [{\citenamefont {Sheng}\ \emph {et~al.}(2017)\citenamefont {Sheng},
  \citenamefont {Perry}, \citenamefont {Krzyzewski}, \citenamefont {Geller},
  \citenamefont {Kitching},\ and\ \citenamefont {Knappe}}]{sheng17}%
  \BibitemOpen
  \bibfield  {author} {\bibinfo {author} {\bibfnamefont {D.}~\bibnamefont
  {Sheng}}, \bibinfo {author} {\bibfnamefont {A.~R.}\ \bibnamefont {Perry}},
  \bibinfo {author} {\bibfnamefont {S.~P.}\ \bibnamefont {Krzyzewski}},
  \bibinfo {author} {\bibfnamefont {S.}~\bibnamefont {Geller}}, \bibinfo
  {author} {\bibfnamefont {J.}~\bibnamefont {Kitching}},\ and\ \bibinfo
  {author} {\bibfnamefont {S.}~\bibnamefont {Knappe}},\ }\bibfield  {title}
  {\bibinfo {title} {A microfabricated optically-pumped magnetic gradiometer},\
  }\href {https://doi.org/10.1063/1.4974349} {\bibfield  {journal} {\bibinfo
  {journal} {Appl. Phys. Lett.}\ }\textbf {\bibinfo {volume} {110}},\ \bibinfo
  {pages} {031106} (\bibinfo {year} {2017})}\BibitemShut {NoStop}%
\bibitem [{\citenamefont {Alem}\ \emph {et~al.}(2017)\citenamefont {Alem},
  \citenamefont {Mhaskar}, \citenamefont {Jim\'{e}nez-Mart\'{i}nez},
  \citenamefont {Sheng}, \citenamefont {LeBlanc}, \citenamefont {Trahms},
  \citenamefont {Sander}, \citenamefont {Kitching},\ and\ \citenamefont
  {Knappe}}]{alem2017}%
  \BibitemOpen
  \bibfield  {author} {\bibinfo {author} {\bibfnamefont {O.}~\bibnamefont
  {Alem}}, \bibinfo {author} {\bibfnamefont {R.}~\bibnamefont {Mhaskar}},
  \bibinfo {author} {\bibfnamefont {R.}~\bibnamefont
  {Jim\'{e}nez-Mart\'{i}nez}}, \bibinfo {author} {\bibfnamefont
  {D.}~\bibnamefont {Sheng}}, \bibinfo {author} {\bibfnamefont
  {J.}~\bibnamefont {LeBlanc}}, \bibinfo {author} {\bibfnamefont
  {L.}~\bibnamefont {Trahms}}, \bibinfo {author} {\bibfnamefont
  {T.}~\bibnamefont {Sander}}, \bibinfo {author} {\bibfnamefont
  {J.}~\bibnamefont {Kitching}},\ and\ \bibinfo {author} {\bibfnamefont
  {S.}~\bibnamefont {Knappe}},\ }\bibfield  {title} {\bibinfo {title} {Magnetic
  field imaging with microfabricated optically-pumped magnetometers},\ }\href
  {https://doi.org/10.1364/OE.25.007849} {\bibfield  {journal} {\bibinfo
  {journal} {Opt. Express}\ }\textbf {\bibinfo {volume} {25}},\ \bibinfo
  {pages} {7849} (\bibinfo {year} {2017})}\BibitemShut {NoStop}%
\bibitem [{\citenamefont {Sulai}\ \emph {et~al.}(2019)\citenamefont {Sulai},
  \citenamefont {DeLand}, \citenamefont {Bulatowicz}, \citenamefont {Wahl},
  \citenamefont {Wakai},\ and\ \citenamefont {Walker}}]{sulai2019}%
  \BibitemOpen
  \bibfield  {author} {\bibinfo {author} {\bibfnamefont {I.~A.}\ \bibnamefont
  {Sulai}}, \bibinfo {author} {\bibfnamefont {Z.~J.}\ \bibnamefont {DeLand}},
  \bibinfo {author} {\bibfnamefont {M.~D.}\ \bibnamefont {Bulatowicz}},
  \bibinfo {author} {\bibfnamefont {C.~P.}\ \bibnamefont {Wahl}}, \bibinfo
  {author} {\bibfnamefont {R.~T.}\ \bibnamefont {Wakai}},\ and\ \bibinfo
  {author} {\bibfnamefont {T.~G.}\ \bibnamefont {Walker}},\ }\bibfield  {title}
  {\bibinfo {title} {Characterizing atomic magnetic gradiometers for fetal
  magnetocardiography},\ }\href {https://doi.org/10.1063/1.5091007} {\bibfield
  {journal} {\bibinfo  {journal} {Review of Scientific Instruments}\ }\textbf
  {\bibinfo {volume} {90}},\ \bibinfo {pages} {085003} (\bibinfo {year}
  {2019})}\BibitemShut {NoStop}%
\bibitem [{\citenamefont {Pratt}\ \emph {et~al.}(2021)\citenamefont {Pratt},
  \citenamefont {Ledbetter}, \citenamefont {Jim{\'e}nez-Mart{\'i}nez},
  \citenamefont {Shapiro}, \citenamefont {Solon}, \citenamefont {Iwata},
  \citenamefont {Garber}, \citenamefont {Gormley}, \citenamefont {Decker},
  \citenamefont {Delgadillo}, \citenamefont {Dellis}, \citenamefont {Phillips},
  \citenamefont {Sundar}, \citenamefont {Leung}, \citenamefont {Coyne},
  \citenamefont {McKinley}, \citenamefont {Lopez}, \citenamefont {Homan},
  \citenamefont {Marsh}, \citenamefont {Zhang}, \citenamefont {Maurice},
  \citenamefont {Siepser}, \citenamefont {Giovannoli}, \citenamefont
  {Leverett}, \citenamefont {Lerner}, \citenamefont {Seidman}, \citenamefont
  {DeLuna}, \citenamefont {Wright-Freeman}, \citenamefont {Kates-Harbeck},
  \citenamefont {Lasser}, \citenamefont {Mohseni}, \citenamefont {Sharp},
  \citenamefont {Zorzos}, \citenamefont {Lara}, \citenamefont {Kouhzadi},
  \citenamefont {Ojeda}, \citenamefont {Chopra}, \citenamefont {Bednarke},
  \citenamefont {Henninger},\ and\ \citenamefont {Alford}}]{kernel2021}%
  \BibitemOpen
  \bibfield  {author} {\bibinfo {author} {\bibfnamefont {E.~J.}\ \bibnamefont
  {Pratt}}, \bibinfo {author} {\bibfnamefont {M.}~\bibnamefont {Ledbetter}},
  \bibinfo {author} {\bibfnamefont {R.}~\bibnamefont
  {Jim{\'e}nez-Mart{\'i}nez}}, \bibinfo {author} {\bibfnamefont
  {B.}~\bibnamefont {Shapiro}}, \bibinfo {author} {\bibfnamefont
  {A.}~\bibnamefont {Solon}}, \bibinfo {author} {\bibfnamefont {G.~Z.}\
  \bibnamefont {Iwata}}, \bibinfo {author} {\bibfnamefont {S.}~\bibnamefont
  {Garber}}, \bibinfo {author} {\bibfnamefont {J.}~\bibnamefont {Gormley}},
  \bibinfo {author} {\bibfnamefont {D.}~\bibnamefont {Decker}}, \bibinfo
  {author} {\bibfnamefont {D.}~\bibnamefont {Delgadillo}}, \bibinfo {author}
  {\bibfnamefont {A.~T.}\ \bibnamefont {Dellis}}, \bibinfo {author}
  {\bibfnamefont {J.}~\bibnamefont {Phillips}}, \bibinfo {author}
  {\bibfnamefont {G.}~\bibnamefont {Sundar}}, \bibinfo {author} {\bibfnamefont
  {J.}~\bibnamefont {Leung}}, \bibinfo {author} {\bibfnamefont
  {J.}~\bibnamefont {Coyne}}, \bibinfo {author} {\bibfnamefont
  {M.}~\bibnamefont {McKinley}}, \bibinfo {author} {\bibfnamefont
  {G.}~\bibnamefont {Lopez}}, \bibinfo {author} {\bibfnamefont
  {S.}~\bibnamefont {Homan}}, \bibinfo {author} {\bibfnamefont
  {L.}~\bibnamefont {Marsh}}, \bibinfo {author} {\bibfnamefont
  {M.}~\bibnamefont {Zhang}}, \bibinfo {author} {\bibfnamefont
  {V.}~\bibnamefont {Maurice}}, \bibinfo {author} {\bibfnamefont
  {B.}~\bibnamefont {Siepser}}, \bibinfo {author} {\bibfnamefont
  {T.}~\bibnamefont {Giovannoli}}, \bibinfo {author} {\bibfnamefont
  {B.}~\bibnamefont {Leverett}}, \bibinfo {author} {\bibfnamefont
  {G.}~\bibnamefont {Lerner}}, \bibinfo {author} {\bibfnamefont
  {S.}~\bibnamefont {Seidman}}, \bibinfo {author} {\bibfnamefont
  {V.}~\bibnamefont {DeLuna}}, \bibinfo {author} {\bibfnamefont
  {K.}~\bibnamefont {Wright-Freeman}}, \bibinfo {author} {\bibfnamefont
  {J.}~\bibnamefont {Kates-Harbeck}}, \bibinfo {author} {\bibfnamefont
  {T.}~\bibnamefont {Lasser}}, \bibinfo {author} {\bibfnamefont
  {H.}~\bibnamefont {Mohseni}}, \bibinfo {author} {\bibfnamefont
  {T.}~\bibnamefont {Sharp}}, \bibinfo {author} {\bibfnamefont
  {A.}~\bibnamefont {Zorzos}}, \bibinfo {author} {\bibfnamefont {A.~H.}\
  \bibnamefont {Lara}}, \bibinfo {author} {\bibfnamefont {A.}~\bibnamefont
  {Kouhzadi}}, \bibinfo {author} {\bibfnamefont {A.}~\bibnamefont {Ojeda}},
  \bibinfo {author} {\bibfnamefont {P.}~\bibnamefont {Chopra}}, \bibinfo
  {author} {\bibfnamefont {Z.}~\bibnamefont {Bednarke}}, \bibinfo {author}
  {\bibfnamefont {M.}~\bibnamefont {Henninger}},\ and\ \bibinfo {author}
  {\bibfnamefont {J.~K.}\ \bibnamefont {Alford}},\ }\bibfield  {title}
  {\bibinfo {title} {{Kernel Flux: a whole-head 432-magnetometer
  optically-pumped magnetoencephalography (OP-MEG) system for brain activity
  imaging during natural human experiences}},\ }in\ \href
  {https://doi.org/10.1117/12.2581794} {\emph {\bibinfo {booktitle} {Optical
  and Quantum Sensing and Precision Metrology}}},\ Vol.\ \bibinfo {volume}
  {11700},\ \bibinfo {editor} {edited by\ \bibinfo {editor} {\bibfnamefont
  {S.~M.}\ \bibnamefont {Shahriar}}\ and\ \bibinfo {editor} {\bibfnamefont
  {J.}~\bibnamefont {Scheuer}}},\ \bibinfo {organization} {International
  Society for Optics and Photonics}\ (\bibinfo  {publisher} {SPIE},\ \bibinfo
  {year} {2021})\ p.\ \bibinfo {pages} {1170032}\BibitemShut {NoStop}%
\bibitem [{\citenamefont {Lucivero}\ \emph {et~al.}(2021)\citenamefont
  {Lucivero}, \citenamefont {Lee}, \citenamefont {Dural},\ and\ \citenamefont
  {Romalis}}]{lucivero2021}%
  \BibitemOpen
  \bibfield  {author} {\bibinfo {author} {\bibfnamefont {V.}~\bibnamefont
  {Lucivero}}, \bibinfo {author} {\bibfnamefont {W.}~\bibnamefont {Lee}},
  \bibinfo {author} {\bibfnamefont {N.}~\bibnamefont {Dural}},\ and\ \bibinfo
  {author} {\bibfnamefont {M.}~\bibnamefont {Romalis}},\ }\bibfield  {title}
  {\bibinfo {title} {Femtotesla direct magnetic gradiometer using a single
  multipass cell},\ }\href {https://doi.org/10.1103/PhysRevApplied.15.014004}
  {\bibfield  {journal} {\bibinfo  {journal} {Phys. Rev. Applied}\ }\textbf
  {\bibinfo {volume} {15}},\ \bibinfo {pages} {014004} (\bibinfo {year}
  {2021})}\BibitemShut {NoStop}%
\bibitem [{\citenamefont {Perry}\ \emph {et~al.}(2020)\citenamefont {Perry},
  \citenamefont {Bulatowicz}, \citenamefont {Larsen}, \citenamefont {Walker},\
  and\ \citenamefont {Wyllie}}]{perry2020}%
  \BibitemOpen
  \bibfield  {author} {\bibinfo {author} {\bibfnamefont {A.~R.}\ \bibnamefont
  {Perry}}, \bibinfo {author} {\bibfnamefont {M.~D.}\ \bibnamefont
  {Bulatowicz}}, \bibinfo {author} {\bibfnamefont {M.}~\bibnamefont {Larsen}},
  \bibinfo {author} {\bibfnamefont {T.~G.}\ \bibnamefont {Walker}},\ and\
  \bibinfo {author} {\bibfnamefont {R.}~\bibnamefont {Wyllie}},\ }\bibfield
  {title} {\bibinfo {title} {All-optical intrinsic atomic gradiometer with
  sub-20 f{T}/cm/$\sqrt{}${H}z sensitivity in a 22 $\mu${T} earth-scale
  magnetic field},\ }\href {https://doi.org/10.1364/OE.408486} {\bibfield
  {journal} {\bibinfo  {journal} {Opt. Express}\ }\textbf {\bibinfo {volume}
  {28}},\ \bibinfo {pages} {36696} (\bibinfo {year} {2020})}\BibitemShut
  {NoStop}%
\bibitem [{\citenamefont {Nabighian}(1972)}]{Nabighian1972}%
  \BibitemOpen
  \bibfield  {author} {\bibinfo {author} {\bibfnamefont {M.~N.}\ \bibnamefont
  {Nabighian}},\ }\bibfield  {title} {\bibinfo {title} {{The analytic signal of
  two-dimensional magnetic bodies with polygonal cross-section; its properties
  and use for automated anomaly interpretation}},\ }\href
  {https://doi.org/10.1190/1.1440276} {\bibfield  {journal} {\bibinfo
  {journal} {Geophysics}\ }\textbf {\bibinfo {volume} {37}},\ \bibinfo {pages}
  {507} (\bibinfo {year} {1972})}\BibitemShut {NoStop}%
\bibitem [{\citenamefont {Nabighian}(1984)}]{Nabighian1984}%
  \BibitemOpen
  \bibfield  {author} {\bibinfo {author} {\bibfnamefont {M.~N.}\ \bibnamefont
  {Nabighian}},\ }\bibfield  {title} {\bibinfo {title} {{Toward a
  three-dimensional automatic interpretation of potential field data via
  generalized Hilbert transforms; fundamental relations}},\ }\href
  {https://doi.org/10.1190/1.1441706} {\bibfield  {journal} {\bibinfo
  {journal} {Geophysics}\ }\textbf {\bibinfo {volume} {49}},\ \bibinfo {pages}
  {780} (\bibinfo {year} {1984})}\BibitemShut {NoStop}%
\bibitem [{\citenamefont {Roest}\ \emph {et~al.}(1992)\citenamefont {Roest},
  \citenamefont {Verhoef},\ and\ \citenamefont {Pilkington}}]{roest1992}%
  \BibitemOpen
  \bibfield  {author} {\bibinfo {author} {\bibfnamefont {W.~R.}\ \bibnamefont
  {Roest}}, \bibinfo {author} {\bibfnamefont {J.}~\bibnamefont {Verhoef}},\
  and\ \bibinfo {author} {\bibfnamefont {M.}~\bibnamefont {Pilkington}},\
  }\bibfield  {title} {\bibinfo {title} {{Magnetic interpretation using the 3-D
  analytic signal}},\ }\href {https://doi.org/10.1190/1.1443174} {\bibfield
  {journal} {\bibinfo  {journal} {Geophysics}\ }\textbf {\bibinfo {volume}
  {57}},\ \bibinfo {pages} {116} (\bibinfo {year} {1992})}\BibitemShut
  {NoStop}%
\bibitem [{\citenamefont {Nelson}(1988)}]{nelson1988}%
  \BibitemOpen
  \bibfield  {author} {\bibinfo {author} {\bibfnamefont {J.~B.}\ \bibnamefont
  {Nelson}},\ }\bibfield  {title} {\bibinfo {title} {{Calculation of the
  magnetic gradient tensor from total field gradient measurements and its
  application to geophysical interpretation}},\ }\href
  {https://doi.org/10.1190/1.1442532} {\bibfield  {journal} {\bibinfo
  {journal} {Geophysics}\ }\textbf {\bibinfo {volume} {53}},\ \bibinfo {pages}
  {957} (\bibinfo {year} {1988})}\BibitemShut {NoStop}%
\bibitem [{\citenamefont {Bell}\ and\ \citenamefont {Bloom}(1961)}]{bell1961}%
  \BibitemOpen
  \bibfield  {author} {\bibinfo {author} {\bibfnamefont {W.~E.}\ \bibnamefont
  {Bell}}\ and\ \bibinfo {author} {\bibfnamefont {A.~L.}\ \bibnamefont
  {Bloom}},\ }\bibfield  {title} {\bibinfo {title} {Optically driven spin
  precession},\ }\href {https://doi.org/10.1103/PhysRevLett.6.280} {\bibfield
  {journal} {\bibinfo  {journal} {Phys. Rev. Lett.}\ }\textbf {\bibinfo
  {volume} {6}},\ \bibinfo {pages} {280} (\bibinfo {year} {1961})}\BibitemShut
  {NoStop}%
\bibitem [{\citenamefont {Appelt}\ \emph {et~al.}(1998)\citenamefont {Appelt},
  \citenamefont {Baranga}, \citenamefont {Erickson}, \citenamefont {Romalis},
  \citenamefont {Young},\ and\ \citenamefont {Happer}}]{appelt98}%
  \BibitemOpen
  \bibfield  {author} {\bibinfo {author} {\bibfnamefont {S.}~\bibnamefont
  {Appelt}}, \bibinfo {author} {\bibfnamefont {A.~B.-A.}\ \bibnamefont
  {Baranga}}, \bibinfo {author} {\bibfnamefont {C.~J.}\ \bibnamefont
  {Erickson}}, \bibinfo {author} {\bibfnamefont {M.~V.}\ \bibnamefont
  {Romalis}}, \bibinfo {author} {\bibfnamefont {A.~R.}\ \bibnamefont {Young}},\
  and\ \bibinfo {author} {\bibfnamefont {W.}~\bibnamefont {Happer}},\
  }\bibfield  {title} {\bibinfo {title} {Theory of spin-exchange optical
  pumping of ${}^{3}\mathrm{He}$ and ${}^{129}\mathrm{Xe}$},\ }\href
  {https://doi.org/10.1103/PhysRevA.58.1412} {\bibfield  {journal} {\bibinfo
  {journal} {Phys. Rev. A}\ }\textbf {\bibinfo {volume} {58}},\ \bibinfo
  {pages} {1412} (\bibinfo {year} {1998})}\BibitemShut {NoStop}%
\bibitem [{\citenamefont {Shah}\ and\ \citenamefont
  {Romalis}(2009)}]{shah2009}%
  \BibitemOpen
  \bibfield  {author} {\bibinfo {author} {\bibfnamefont {V.}~\bibnamefont
  {Shah}}\ and\ \bibinfo {author} {\bibfnamefont {M.~V.}\ \bibnamefont
  {Romalis}},\ }\bibfield  {title} {\bibinfo {title} {Spin-exchange
  relaxation-free magnetometry using elliptically polarized light},\ }\href
  {https://doi.org/10.1103/PhysRevA.80.013416} {\bibfield  {journal} {\bibinfo
  {journal} {Phys. Rev. A}\ }\textbf {\bibinfo {volume} {80}},\ \bibinfo
  {pages} {013416} (\bibinfo {year} {2009})}\BibitemShut {NoStop}%
\bibitem [{\citenamefont {Cai}\ \emph {et~al.}(2020)\citenamefont {Cai},
  \citenamefont {Hao}, \citenamefont {Qiu}, \citenamefont {Yu}, \citenamefont
  {Xiao},\ and\ \citenamefont {Sheng}}]{cai2020}%
  \BibitemOpen
  \bibfield  {author} {\bibinfo {author} {\bibfnamefont {B.}~\bibnamefont
  {Cai}}, \bibinfo {author} {\bibfnamefont {C.-P.}\ \bibnamefont {Hao}},
  \bibinfo {author} {\bibfnamefont {Z.-R.}\ \bibnamefont {Qiu}}, \bibinfo
  {author} {\bibfnamefont {Q.-Q.}\ \bibnamefont {Yu}}, \bibinfo {author}
  {\bibfnamefont {W.}~\bibnamefont {Xiao}},\ and\ \bibinfo {author}
  {\bibfnamefont {D.}~\bibnamefont {Sheng}},\ }\bibfield  {title} {\bibinfo
  {title} {Herriott-cavity-assisted all-optical atomic vector magnetometer},\
  }\href {https://doi.org/10.1103/PhysRevA.101.053436} {\bibfield  {journal}
  {\bibinfo  {journal} {Phys. Rev. A}\ }\textbf {\bibinfo {volume} {101}},\
  \bibinfo {pages} {053436} (\bibinfo {year} {2020})}\BibitemShut {NoStop}%
\bibitem [{\citenamefont {Yu}\ \emph {et~al.}(2022)\citenamefont {Yu},
  \citenamefont {Liu}, \citenamefont {Yuan},\ and\ \citenamefont
  {Sheng}}]{yu2022}%
  \BibitemOpen
  \bibfield  {author} {\bibinfo {author} {\bibfnamefont {Q.-Q.}\ \bibnamefont
  {Yu}}, \bibinfo {author} {\bibfnamefont {S.-Q.}\ \bibnamefont {Liu}},
  \bibinfo {author} {\bibfnamefont {C.-Q.}\ \bibnamefont {Yuan}},\ and\
  \bibinfo {author} {\bibfnamefont {D.}~\bibnamefont {Sheng}},\ }\bibfield
  {title} {\bibinfo {title} {Light-shift-free and dead-zone-free
  atomic-orientation-based scalar magnetometry using a single
  amplitude-modulated beam},\ }\href
  {https://doi.org/10.1103/PhysRevApplied.18.014015} {\bibfield  {journal}
  {\bibinfo  {journal} {Phys. Rev. Applied}\ }\textbf {\bibinfo {volume}
  {18}},\ \bibinfo {pages} {014015} (\bibinfo {year} {2022})}\BibitemShut
  {NoStop}%
\bibitem [{\citenamefont {Happer}(1972)}]{happer72}%
  \BibitemOpen
  \bibfield  {author} {\bibinfo {author} {\bibfnamefont {W.}~\bibnamefont
  {Happer}},\ }\bibfield  {title} {\bibinfo {title} {Optical pumping},\ }\href
  {https://doi.org/10.1103/RevModPhys.44.169} {\bibfield  {journal} {\bibinfo
  {journal} {Rev. Mod. Phys.}\ }\textbf {\bibinfo {volume} {44}},\ \bibinfo
  {pages} {169} (\bibinfo {year} {1972})}\BibitemShut {NoStop}%
\bibitem [{\citenamefont {Jones}(1941)}]{jones1941}%
  \BibitemOpen
  \bibfield  {author} {\bibinfo {author} {\bibfnamefont {R.~C.}\ \bibnamefont
  {Jones}},\ }\bibfield  {title} {\bibinfo {title} {A new calculus for the
  treatment of optical systemsi. description and discussion of the calculus},\
  }\href {https://doi.org/10.1364/JOSA.31.000488} {\bibfield  {journal}
  {\bibinfo  {journal} {J. Opt. Soc. Am.}\ }\textbf {\bibinfo {volume} {31}},\
  \bibinfo {pages} {488} (\bibinfo {year} {1941})}\BibitemShut {NoStop}%
\bibitem [{\citenamefont {Jacobs}(1991)}]{jacobsthesis}%
  \BibitemOpen
  \bibfield  {author} {\bibinfo {author} {\bibfnamefont {J.~P.}\ \bibnamefont
  {Jacobs}},\ }\href
  {http://www.pqdtcn.com/thesisDetails/0C0E219DCAAFBCA260D5127DE4F70917}
  {\bibinfo {type} {Ph.d. thesis}},\ \bibinfo  {school} {University of
  Washington} (\bibinfo {year} {1991})\BibitemShut {NoStop}%
\bibitem [{\citenamefont {Li}\ \emph {et~al.}(2011)\citenamefont {Li},
  \citenamefont {Vachaspati}, \citenamefont {Sheng}, \citenamefont {Dural},\
  and\ \citenamefont {Romalis}}]{li2011}%
  \BibitemOpen
  \bibfield  {author} {\bibinfo {author} {\bibfnamefont {S.}~\bibnamefont
  {Li}}, \bibinfo {author} {\bibfnamefont {P.}~\bibnamefont {Vachaspati}},
  \bibinfo {author} {\bibfnamefont {D.}~\bibnamefont {Sheng}}, \bibinfo
  {author} {\bibfnamefont {N.}~\bibnamefont {Dural}},\ and\ \bibinfo {author}
  {\bibfnamefont {M.~V.}\ \bibnamefont {Romalis}},\ }\bibfield  {title}
  {\bibinfo {title} {Optical rotation in excess of 100 rad generated by rb
  vapor in a multipass cell},\ }\href
  {https://doi.org/10.1103/PhysRevA.84.061403} {\bibfield  {journal} {\bibinfo
  {journal} {Phys. Rev. A}\ }\textbf {\bibinfo {volume} {84}},\ \bibinfo
  {pages} {061403} (\bibinfo {year} {2011})}\BibitemShut {NoStop}%
\bibitem [{\citenamefont {Romalis}\ \emph {et~al.}(1997)\citenamefont
  {Romalis}, \citenamefont {Miron},\ and\ \citenamefont {Cates}}]{romalis1997}%
  \BibitemOpen
  \bibfield  {author} {\bibinfo {author} {\bibfnamefont {M.~V.}\ \bibnamefont
  {Romalis}}, \bibinfo {author} {\bibfnamefont {E.}~\bibnamefont {Miron}},\
  and\ \bibinfo {author} {\bibfnamefont {G.~D.}\ \bibnamefont {Cates}},\
  }\bibfield  {title} {\bibinfo {title} {Pressure broadening of rb {D}$_{1}$
  and {D}$_{2}$ lines by ${}^{3}${He}, ${}^{4}${He}, {N}${}_{2}$, and {Xe}:
  Line cores and near wings},\ }\href
  {https://doi.org/10.1103/PhysRevA.56.4569} {\bibfield  {journal} {\bibinfo
  {journal} {Phys. Rev. A}\ }\textbf {\bibinfo {volume} {56}},\ \bibinfo
  {pages} {4569} (\bibinfo {year} {1997})}\BibitemShut {NoStop}%
\bibitem [{\citenamefont {Scholtes}\ \emph {et~al.}(2011)\citenamefont
  {Scholtes}, \citenamefont {Schultze}, \citenamefont {IJsselsteijn},
  \citenamefont {Woetzel},\ and\ \citenamefont {Meyer}}]{scholtes2011}%
  \BibitemOpen
  \bibfield  {author} {\bibinfo {author} {\bibfnamefont {T.}~\bibnamefont
  {Scholtes}}, \bibinfo {author} {\bibfnamefont {V.}~\bibnamefont {Schultze}},
  \bibinfo {author} {\bibfnamefont {R.}~\bibnamefont {IJsselsteijn}}, \bibinfo
  {author} {\bibfnamefont {S.}~\bibnamefont {Woetzel}},\ and\ \bibinfo {author}
  {\bibfnamefont {H.-G.}\ \bibnamefont {Meyer}},\ }\bibfield  {title} {\bibinfo
  {title} {Light-narrowed optically pumped ${M}_{x}$ magnetometer with a
  miniaturized {Cs} cell},\ }\href {https://doi.org/10.1103/PhysRevA.84.043416}
  {\bibfield  {journal} {\bibinfo  {journal} {Phys. Rev. A}\ }\textbf {\bibinfo
  {volume} {84}},\ \bibinfo {pages} {043416} (\bibinfo {year}
  {2011})}\BibitemShut {NoStop}%
\bibitem [{\citenamefont {Patton}\ \emph {et~al.}(2014)\citenamefont {Patton},
  \citenamefont {Zhivun}, \citenamefont {Hovde},\ and\ \citenamefont
  {Budker}}]{patton2014}%
  \BibitemOpen
  \bibfield  {author} {\bibinfo {author} {\bibfnamefont {B.}~\bibnamefont
  {Patton}}, \bibinfo {author} {\bibfnamefont {E.}~\bibnamefont {Zhivun}},
  \bibinfo {author} {\bibfnamefont {D.~C.}\ \bibnamefont {Hovde}},\ and\
  \bibinfo {author} {\bibfnamefont {D.}~\bibnamefont {Budker}},\ }\bibfield
  {title} {\bibinfo {title} {All-optical vector atomic magnetometer},\ }\href
  {https://doi.org/10.1103/PhysRevLett.113.013001} {\bibfield  {journal}
  {\bibinfo  {journal} {Phys. Rev. Lett.}\ }\textbf {\bibinfo {volume} {113}},\
  \bibinfo {pages} {013001} (\bibinfo {year} {2014})}\BibitemShut {NoStop}%
\bibitem [{\citenamefont {Zheng}\ \emph {et~al.}(2020)\citenamefont {Zheng},
  \citenamefont {Sun}, \citenamefont {Chatzidrosos}, \citenamefont {Zhang},
  \citenamefont {Nakamura}, \citenamefont {Sumiya}, \citenamefont {Ohshima},
  \citenamefont {Isoya}, \citenamefont {Wrachtrup}, \citenamefont
  {Wickenbrock},\ and\ \citenamefont {Budker}}]{zheng2020}%
  \BibitemOpen
  \bibfield  {author} {\bibinfo {author} {\bibfnamefont {H.}~\bibnamefont
  {Zheng}}, \bibinfo {author} {\bibfnamefont {Z.}~\bibnamefont {Sun}}, \bibinfo
  {author} {\bibfnamefont {G.}~\bibnamefont {Chatzidrosos}}, \bibinfo {author}
  {\bibfnamefont {C.}~\bibnamefont {Zhang}}, \bibinfo {author} {\bibfnamefont
  {K.}~\bibnamefont {Nakamura}}, \bibinfo {author} {\bibfnamefont
  {H.}~\bibnamefont {Sumiya}}, \bibinfo {author} {\bibfnamefont
  {T.}~\bibnamefont {Ohshima}}, \bibinfo {author} {\bibfnamefont
  {J.}~\bibnamefont {Isoya}}, \bibinfo {author} {\bibfnamefont
  {J.}~\bibnamefont {Wrachtrup}}, \bibinfo {author} {\bibfnamefont
  {A.}~\bibnamefont {Wickenbrock}},\ and\ \bibinfo {author} {\bibfnamefont
  {D.}~\bibnamefont {Budker}},\ }\bibfield  {title} {\bibinfo {title}
  {Microwave-free vector magnetometry with nitrogen-vacancy centers along a
  single axis in diamond},\ }\href
  {https://doi.org/10.1103/PhysRevApplied.13.044023} {\bibfield  {journal}
  {\bibinfo  {journal} {Phys. Rev. Appl.}\ }\textbf {\bibinfo {volume} {13}},\
  \bibinfo {pages} {044023} (\bibinfo {year} {2020})}\BibitemShut {NoStop}%
\bibitem [{\citenamefont {Wang}\ \emph {et~al.}(2022)\citenamefont {Wang},
  \citenamefont {Peng}, \citenamefont {Xiao},\ and\ \citenamefont
  {Guo}}]{wang2022}%
  \BibitemOpen
  \bibfield  {author} {\bibinfo {author} {\bibfnamefont {B.}~\bibnamefont
  {Wang}}, \bibinfo {author} {\bibfnamefont {X.}~\bibnamefont {Peng}}, \bibinfo
  {author} {\bibfnamefont {W.}~\bibnamefont {Xiao}},\ and\ \bibinfo {author}
  {\bibfnamefont {H.}~\bibnamefont {Guo}},\ }\bibfield  {title} {\bibinfo
  {title} {Three-axis isotropic-sensitivity $^{4}\mathrm{He}$ magnetometer with
  alignment-based longitudinal parametric resonance},\ }\href
  {https://doi.org/10.1103/PhysRevA.106.063102} {\bibfield  {journal} {\bibinfo
   {journal} {Phys. Rev. A}\ }\textbf {\bibinfo {volume} {106}},\ \bibinfo
  {pages} {063102} (\bibinfo {year} {2022})}\BibitemShut {NoStop}%
\bibitem [{\citenamefont {Lee}\ \emph {et~al.}(2021)\citenamefont {Lee},
  \citenamefont {Lucivero}, \citenamefont {Romalis}, \citenamefont {Limes},
  \citenamefont {Foley},\ and\ \citenamefont {Kornack}}]{lee2021}%
  \BibitemOpen
  \bibfield  {author} {\bibinfo {author} {\bibfnamefont {W.}~\bibnamefont
  {Lee}}, \bibinfo {author} {\bibfnamefont {V.~G.}\ \bibnamefont {Lucivero}},
  \bibinfo {author} {\bibfnamefont {M.~V.}\ \bibnamefont {Romalis}}, \bibinfo
  {author} {\bibfnamefont {M.~E.}\ \bibnamefont {Limes}}, \bibinfo {author}
  {\bibfnamefont {E.~L.}\ \bibnamefont {Foley}},\ and\ \bibinfo {author}
  {\bibfnamefont {T.~W.}\ \bibnamefont {Kornack}},\ }\bibfield  {title}
  {\bibinfo {title} {Heading errors in all-optical alkali-metal-vapor
  magnetometers in geomagnetic fields},\ }\href
  {https://doi.org/10.1103/PhysRevA.103.063103} {\bibfield  {journal} {\bibinfo
   {journal} {Phys. Rev. A}\ }\textbf {\bibinfo {volume} {103}},\ \bibinfo
  {pages} {063103} (\bibinfo {year} {2021})}\BibitemShut {NoStop}%
\end{thebibliography}
\end{document}